\documentclass[lettersize,journal]{IEEEtran}
\usepackage{amssymb,amsfonts}
\usepackage{algorithmic}
\usepackage{graphicx}
\usepackage{textcomp}
\usepackage{amsmath}
\usepackage{bm}
\usepackage{empheq}
\usepackage[font=small,labelfont=bf]{caption}
\usepackage{subcaption}
\usepackage[normalem]{ulem}
\usepackage{cancel}
\usepackage{makecell}
\usepackage{amssymb}
\usepackage{rotating}
\usepackage{booktabs}
\usepackage{longtable}
\usepackage{multirow}
\usepackage{tabularx}
\usepackage{adjustbox}
\usepackage{cite}
\def\BibTeX{{\rm B\kern-.05em{\sc i\kern-.025em b}\kern-.08em
    T\kern-.1667em\lower.7ex\hbox{E}\kern-.125emX}}
    
\usepackage{balance}
\usepackage{url}
\usepackage{hyperref}

\begin{document}

\title{Rainfall Sensing via Mobile Communication Signals}
\author{Zhongqin Wang, \IEEEmembership{Member, IEEE},
		J. Andrew Zhang, \IEEEmembership{Senior Member, IEEE},\\
		Kai Wu, \IEEEmembership{Member, IEEE},
		Y. Jay Guo, \IEEEmembership{Life Fellow, IEEE}

\IEEEcompsocitemizethanks{
\IEEEcompsocthanksitem Zhongqin Wang, J. Andrew Zhang (Corresponding Author), Kai Wu, and Y. Jay Guo are with the Global Big Data Technologies Centre, University of Technology Sydney, Sydney 2007, Australia. E-mail: \{zhongqin.wang, andrew.zhang, kai.wu\}@uts.edu.au; jay.guo@uts.edu.au
}}
\maketitle

\begin{abstract}
Rainfall monitoring is important for hydrological observation, disaster warning, and environmental sensing, but conventional rain gauges and weather radars suffer from sparse deployment and high infrastructure costs. This paper proposes \textit{PMN-RainSense}, a rainfall sensing framework using sub-6-GHz mobile communication signals that supports practical single-antenna deployment. Unlike attenuation-based approaches, which are unreliable at sub-6~GHz because rain-induced attenuation over short mobile access links is only on the order of hundredths of a decibel, the proposed framework exploits fine-grained dynamics. A spectral--temporal channel state information (CSI) compensation method suppresses packet-wise timing and phase distortions while preserving sensing-relevant information. Rainfall-sensitive features are extracted from the delay--Doppler domain to mitigate environmental interference, with angle-domain filtering as an optional extension for multi-antenna receivers. Under bandwidth and antenna constraints, rainfall-correlated Doppler fluctuations serve as the dominant sensing signature, while Doppler-domain normalization improves robustness across links and deployments. Controlled WiFi experiments demonstrate rainfall-associated Doppler broadening and achieve a three-class classification accuracy of 95.48\% using a random forest classifier. Long-Term Evolution (LTE) CSI measurements collected from cellular base stations over 11 carrier frequencies from 0.763 to 2.68~GHz yield a mean absolute error (MAE) of 0.25--0.27~mm/h for rainfall intensity estimation using a one-dimensional convolutional network.
\end{abstract}


\section{Introduction}
Accurate rainfall monitoring is essential for flood warning, urban drainage management, and transportation safety. Rainfall exhibits strong spatial and temporal variability, particularly during intense events, when precipitation can differ substantially across nearby areas. In urban environments, such localized rainfall can rapidly overload drainage systems and trigger flooding, motivating dense and continuous monitoring. Conventional approaches mainly rely on rain gauges \cite{wang2023traditional}, weather radars \cite{borga2022rainfall, oydvin2025combining}, and satellite observations \cite{karthika2026satellite}. Rain gauges are accurate but sparsely deployed and costly to densify. Weather radars offer wide-area coverage but require dedicated infrastructure and may suffer from blockage, signal attenuation, and limited near-surface accuracy. Satellite systems provide large-scale observations but have limited temporal resolution and local sensitivity. Consequently, existing systems face a tradeoff among accuracy, spatial density, coverage, and cost, complicating fine-grained rainfall monitoring in practice.

Recent advances in perceptive mobile networks (PMNs) \cite{zhang2021overview, 11231721} and integrated sensing and communications (ISAC) \cite{lu2024integrated} have created new opportunities for opportunistic environmental sensing using existing mobile infrastructure \cite{wu2025ISAC}. Because mobile networks are already widely deployed, their signals continuously interact with the surrounding environment during propagation, providing a low-cost and pervasive alternative to dedicated sensing infrastructure. Existing communication-signal-based rainfall sensing mainly relies on attenuation or coarse radio measurements, including received signal strength indicator (RSSI), reference signal received power (RSRP), and reference signal received quality (RSRQ) reported by Long-Term Evolution (LTE) user equipment (UE) \cite{wu2025ISAC, 11322988, 11151998, 11162390, xu2025smartphone}, as well as attenuation statistics from commercial microwave links (CMLs) \cite{overeem2011measuring, overeem2016retrieval, nielsen2024merging, 11099507}. Rainfall intensity is then inferred from received-power variations or statistical relationships between these coarse measurements and precipitation.

However, rain-induced attenuation is extremely weak over sub-6-GHz mobile access links. According to the ITU-R model \cite{itu838}, even at 2.6~GHz under heavy rainfall of approximately 50~mm/h, the specific attenuation is on the order of $10^{-2}$~dB/km. Since urban access links typically span several hundred meters to approximately 1~km, the resulting received-power variation is generally limited to a few hundredths of a decibel or less. Detecting such weak variations is difficult in practical networks, where the RSSI, RSRP, and RSRQ are further affected by automatic gain control (AGC), transmit-power adaptation, scheduling, interference, quantization, and hardware-dependent uncertainty. CML-based methods remain effective because these links are longer, more directional, and usually operate at higher frequencies, whereas sub-6-GHz access links exhibit much weaker attenuation responses and more complex channel dynamics. Consequently, coarse received-power measurements provide limited sensitivity to rainfall, while the fine-grained temporal, multipath, and spatial channel variations induced by raindrops remain largely unexploited.

\begin{figure*}[!t]
\centering
\includegraphics[width=\textwidth]{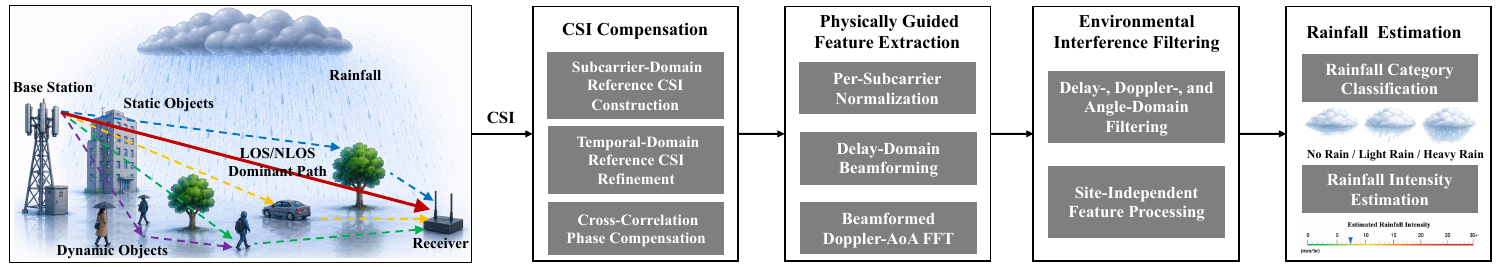}
\caption{Overview of the proposed bistatic CSI-based rainfall sensing framework using sub-6 GHz mobile communication signals.}
\label{fig:overview}
\vspace{-1em}
\end{figure*}

Compared with coarse received-power indicators, channel state information (CSI) preserves subcarrier-level amplitude and phase across time and antennas and has been widely exploited for wireless sensing applications, including localization and tracking \cite{wang2023single, pegoraro2024jump, 10737138, 11498405, wang2026rethinking}, and water-level sensing \cite{11432153, 11605966}. Raindrops absorb and scatter electromagnetic waves, attenuating dominant propagation paths and introducing random phase perturbations, while the distributed raindrop field produces weak dynamic scattering \cite{nia2025exploring}. These effects jointly perturb the delay, Doppler, and spatial characteristics of the channel, altering its short-term structure and reducing temporal coherence. They therefore appear as fine-grained fluctuations in CSI amplitude and phase. Unlike path attenuation, which is a first-order power effect accumulated along the propagation path, these fluctuations arise from time-varying channel dynamics and are not determined solely by link distance. CSI-based sensing can capture rainfall-induced perturbations that coarse signal-strength indicators cannot resolve.

Despite this potential, rainfall sensing using raw CSI remains challenging. First, in practical bistatic systems, CSI is corrupted by clock asynchronization and hardware impairments \cite{zhang2022integration}, including AGC-induced amplitude scaling, timing-offset (TO)-induced phase distortion across subcarriers, residual carrier-frequency-offset (CFO)-induced common phase rotation, and antenna-dependent phase offsets (POs). These distortions may vary across CSI snapshots, introducing artificial fluctuations unrelated to rainfall. Second, rainfall-induced CSI perturbations are inherently weak at sub-6~GHz because the longer wavelength reduces sensitivity to raindrop scattering, while attenuation accumulates over only short access-link distances. The resulting power and multipath variations can therefore be easily overwhelmed by environmental dynamics, including traffic, pedestrian motion, and vegetation movement. Reliable rainfall sensing consequently requires physics-guided CSI preprocessing and feature extraction to suppress measurement distortions, enhance rainfall-sensitive Doppler fluctuations, and mitigate environmental interference.

To address these challenges, this work proposes \textit{PMN-RainSense}, a bistatic CSI-based rainfall sensing framework, as illustrated in Fig.~\ref{fig:overview}. The framework supports a single receive antenna and can be extended to multi-antenna receivers. We first establish a bistatic CSI model that jointly characterizes rainfall-affected propagation and practical CSI impairments. Guided by this model, packet-wise reference CSI is constructed across the subcarrier and temporal domains to suppress timing- and phase-related distortions. The compensated CSI is transformed into delay--Doppler features, with optional angle-domain refinement when multiple receive antennas are available. Under practical bandwidth and antenna-aperture constraints, Doppler processing provides greater sensitivity to rainfall-induced channel dynamics than the delay and angle domains. We therefore adopt rainfall-sensitive Doppler fluctuations as the dominant sensing signature. To the best of our knowledge, this is the first work to exploit Doppler-domain features from bistatic sub-6-GHz CSI for rainfall sensing.

Our main contributions are summarized as follows:
\begin{itemize}
    \item We establish a bistatic CSI rainfall sensing model that jointly characterizes baseline propagation, rainfall-induced perturbations, and CSI impairments, providing a physical foundation for extracting rainfall-sensitive features from sub-6-GHz mobile communication signals.

    \item We propose a packet-wise CSI compensation framework based on reference construction across the subcarrier and temporal domains. The method suppresses timing- and phase-related distortions using a single receive antenna, without requiring antenna pairing or hardware.

    \item We identify rainfall-correlated Doppler fluctuations as the dominant sensing signature in bistatic CSI and develop a physics-guided delay--Doppler feature extraction and normalization scheme to mitigate environmental interference and improve robustness across links and deployments.

    \item We validate \textit{PMN-RainSense} using controlled WiFi measurements and long-term outdoor LTE measurements. The framework achieves a three-class rainfall classification accuracy of 95.48\% and a rainfall-intensity MAE of 0.25--0.27~mm/h on a held-out LTE link at 2.68~GHz.
\end{itemize}

\section{Bistatic CSI Modeling for Rainfall Sensing}
We consider a bistatic communication system representative of practical mobile-network deployments, where the transmitter and receiver are spatially separated, fixed in position, and lack accurate clock synchronization. The transmitter is equipped with a single antenna, while the receiver employs an $M$-element antenna array. Let $N$ and $L$ denote the numbers of subcarriers and CSI snapshots within one CPI, respectively. The CSI measured at the $i$-th subcarrier, the $m$-th receive antenna, and the $k$-th snapshot is denoted by $C_{i,m,k}$, where $i=1,\ldots,N$, $m=1,\ldots,M$, and $k=1,\ldots,L$. For snapshot $k$, the CSI samples across all subcarriers and receive antennas form the matrix $\mathbf{C}_k\in\mathbb{C}^{N\times M}$, whose $(i,m)$-th entry is $C_{i,m,k}$. The measured CSI is modeled as
\begin{equation}
\mathbf{C}_k
=
g_k
e^{-j\varphi^{\mathrm{CFO}}_k}
\mathbf{D}^{\mathrm{TO}}_k
\left(
\mathbf{H}^{\mathrm{B}}_k
+
\mathbf{H}^{\mathrm{X}}_k
+
\mathbf{H}^{\mathrm{R}}_k
\right)
\mathbf{D}^{\mathrm{PO}}_k,
\label{eq:C_model_1}
\end{equation}
where $j=\sqrt{-1}$ is the imaginary unit. The terms $g_k$, $\varphi^{\mathrm{CFO}}_k$, $\mathbf{D}^{\mathrm{TO}}_k$, and $\mathbf{D}^{\mathrm{PO}}_k$ are snapshot-dependent CSI impairments while $\mathbf{H}^{\mathrm{B}}_k$, $\mathbf{H}^{\mathrm{X}}_k$, and $\mathbf{H}^{\mathrm{R}}_k$ describe the baseline propagation component, environmental dynamic component, and rainfall-induced scattering component. These terms are detailed below.

\subsubsection{AGC}
AGC maintains the received signal within the dynamic range of the analog-to-digital converter. Its effect, together with other common receiver-side amplitude scaling, is represented by the positive snapshot-dependent scalar $g_k$ in Eq.~\eqref{eq:C_model_1}. The scalar $g_k$ captures snapshot-dependent common amplitude scaling across subcarriers and receive antennas, which may introduce artificial CSI fluctuations unrelated to rainfall. Its effect can be mitigated by calibrating each CSI snapshot using a receiver-reported power reference \cite{halperin2011tool, wei2022rssi}. \textit{Hereafter, the CSI is assumed to have been compensated for AGC-induced amplitude scaling.}

\subsubsection{TO}
TO introduces a phase rotation that varies linearly with subcarrier frequency. It is modeled as
\begin{equation}
\mathbf{D}^{\mathrm{TO}}_k
=
\mathrm{diag}
\!\left(
e^{-j2\pi f_1 \tau^{\mathrm{TO}}_k},
\ldots,
e^{-j2\pi f_N \tau^{\mathrm{TO}}_k}
\right),
\label{eq:D_to}
\end{equation}
where $f_i$ is the $i$-th subcarrier frequency offset, and $\tau^{\mathrm{TO}}_k$ is the TO at snapshot $k$. Thus, TO produces a frequency-dependent phase distortion across subcarriers.

\subsubsection{CFO}
The residual CFO $\varphi^{\mathrm{CFO}}_k$ comes from the frequency mismatch between the transmitter and receiver oscillators. This rotation is common to all subcarriers and receive antennas within a snapshot but varies across snapshots.

\subsubsection{PO}
PO captures antenna-dependent phase distortions caused by phase-locked loop initialization uncertainty \cite{zubow2021phase} and receive-chain responses \cite{wang2023single}, which may differ across receive antennas and vary across CSI snapshots, as represented by
\begin{equation}
\mathbf{D}^{\mathrm{PO}}_k
=
\mathrm{diag}
\!\left(
e^{-j\phi^{\mathrm{PO}}_{1,k}},
\ldots,
e^{-j\phi^{\mathrm{PO}}_{M,k}}
\right),
\label{eq:D_po}
\end{equation}
where $\phi^{\mathrm{PO}}_{m,k}$ denotes the PO associated with the $m$-th receive antenna at snapshot $k$.

\subsubsection{Baseline Propagation Component}
The baseline propagation component includes dominant line-of-sight (LoS) and non-line-of-sight (NLoS) paths, together with reflections from static objects in the environment. It is modeled as
\begin{equation}
\mathbf{H}^{\mathrm{B}}_k
=
\sum_{\ell=1}^{\mathcal{L}_{\mathrm{B}}}
\alpha^{\mathrm{B}}_{\ell,k}
\mathbf{b}\!\left(\tau^{\mathrm{B}}_{\ell}\right)
\mathbf{a}^{H}\!\left(\theta^{\mathrm{B}}_{\ell}\right),
\label{eq:H_baseline}
\end{equation}
where $\mathcal{L}_{\mathrm{B}}$ is the number of baseline paths, while $\tau^{\mathrm{B}}_{\ell}$ and $\theta^{\mathrm{B}}_{\ell}$ are the propagation delay and angle of arrival (AoA) of the $\ell$-th path, respectively. The corresponding complex gain is
\begin{equation}
\alpha^{\mathrm{B}}_{\ell,k}
=
\rho^{\mathrm{B}}_{\ell}
\epsilon^{\mathrm{B}}_{\ell,k}
e^{-j\psi^{\mathrm{B}}_{\ell,k}},
\label{eq:alpha_baseline}
\end{equation}
where $\rho^{\mathrm{B}}_{\ell}$ is the no-rain complex gain, $\epsilon^{\mathrm{B}}_{\ell,k}$ is the rainfall-induced amplitude scaling, and $\psi^{\mathrm{B}}_{\ell,k}$ is the rainfall-induced phase perturbation. Under no-rain conditions, $\epsilon^{\mathrm{B}}_{\ell,k}=1$ and $\psi^{\mathrm{B}}_{\ell,k}=0$, such that $\alpha^{\mathrm{B}}_{\ell,k}=\rho^{\mathrm{B}}_{\ell}$ and the baseline component is approximately time-invariant. The frequency-domain steering vector is defined as
\begin{equation}
\mathbf{b}\!\left(\tau\right)
=
\left[
e^{-j2\pi f_1 \tau},
\ldots,
e^{-j2\pi f_N \tau}
\right]^T
\in \mathbb{C}^{N \times 1},
\label{eq:delay_vector}
\end{equation}
The receive-array steering vector is
\begin{equation}
\mathbf{a}\!\left(\theta \right)
=
\left[
1,
e^{-j2\pi \frac{d\sin\theta}{\lambda}},
\ldots,
e^{-j2\pi (M-1)\frac{d\sin\theta}{\lambda}}
\right]^T
\in \mathbb{C}^{M \times 1},
\label{eq:angle_vector}
\end{equation}
where $d$ is the antenna spacing and $\lambda$ is the carrier wavelength.

\subsubsection{Environmental Dynamic Component}
The environmental dynamic component represents time-varying propagation paths generated by non-rain objects, including vehicles, pedestrians, and vegetation. These paths may be perturbed by rainfall through amplitude and phase variations. It is modeled as
\begin{equation}
\mathbf{H}^{\mathrm{X}}_k
=
\sum_{\ell=1}^{\mathcal{L}_{\mathrm{X}}}
\alpha^{\mathrm{X}}_{\ell,k}
e^{-j2\pi f^{\mathrm{X}}_{\mathrm{D},\ell}(k-1)\Delta t}
\mathbf{b}\!\left(\tau^{\mathrm{X}}_{\ell}\right)
\mathbf{a}^{H}\!\left(\theta^{\mathrm{X}}_{\ell}\right),
\label{eq:H_dynamic}
\end{equation}
where $\mathcal{L}_{\mathrm{X}}$ is the number of environmental dynamic paths, while $\tau^{\mathrm{X}}_{\ell}$, $\theta^{\mathrm{X}}_{\ell}$, and $f^{\mathrm{X}}_{\mathrm{D},\ell}$ denote the delay, AoA, and Doppler frequency of the $\ell$-th path, respectively. The snapshot interval is denoted by $\Delta t$. These parameters are assumed to be locally constant within one coherent processing interval (CPI) \cite{zhang2021overview} but may vary across CPIs. The complex gain is
\begin{equation}
\alpha^{\mathrm{X}}_{\ell,k}
=
\rho^{\mathrm{X}}_{\ell}
\epsilon^{\mathrm{X}}_{\ell,k}
e^{-j\psi^{\mathrm{X}}_{\ell,k}},
\label{eq:alpha_dynamic}
\end{equation}
where $\rho^{\mathrm{X}}_{\ell}$ is the no-rain complex gain, and $\epsilon^{\mathrm{X}}_{\ell,k}$ and $\psi^{\mathrm{X}}_{\ell,k}$ represent rainfall-induced amplitude scaling and phase perturbation. Under no-rain conditions, $\epsilon^{\mathrm{X}}_{\ell,k}=1$ and $\psi^{\mathrm{X}}_{\ell,k}=0$.

\subsubsection{Rainfall-Induced Scattering Component}
Besides perturbing existing paths, rainfall introduces weak scattering from distributed raindrops. Their varying locations, sizes, and velocities produce superimposed Doppler responses, modeled using a finite number of scattering components as
\begin{equation}
\mathbf{H}^{\mathrm{R}}_k
=
\sum_{\ell=1}^{\mathcal{L}_{\mathrm{R}}}
\rho^{\mathrm{R}}_{\ell,k}
e^{-j2\pi f^{\mathrm{R}}_{\mathrm{D},\ell}(k-1)\Delta t}
\mathbf{b}\!\left(\tau^{\mathrm{R}}_{\ell}\right)
\mathbf{a}^{H}\!\left(\theta^{\mathrm{R}}_{\ell}\right),
\label{eq:H_rain}
\end{equation}
where $\mathcal{L}_{\mathrm{R}}$ is the number of rain scattering components. The quantities $\rho^{\mathrm{R}}_{\ell,k}$, $\tau^{\mathrm{R}}_{\ell}$, $\theta^{\mathrm{R}}_{\ell}$, and $f^{\mathrm{R}}_{\mathrm{D},\ell}$ denote the complex gain, delay, AoA, and Doppler frequency of the $\ell$-th component, respectively. Following the same local-stationarity assumption, the delay, AoA, and Doppler frequency remain approximately constant within a CPI, while $\rho^{\mathrm{R}}_{\ell,k}$ captures short-term amplitude and phase fluctuations. These parameters may evolve across CPIs as the scattering geometry and raindrop-velocity distribution change. Under no-rain conditions, this component is absent, i.e., $\mathbf{H}^{\mathrm{R}}_k=\mathbf{0}$.

The measured CSI therefore contains rainfall-sensitive channel variations together with snapshot-dependent AGC, TO, CFO, and PO distortions. CSI compensation is consequently required before reliable rainfall-sensitive features can be extracted, as described in the next section.

\section{Joint Spectral-Temporal CSI Compensation}
\begin{figure}[!t]
\centering
\includegraphics[width=0.48\textwidth]{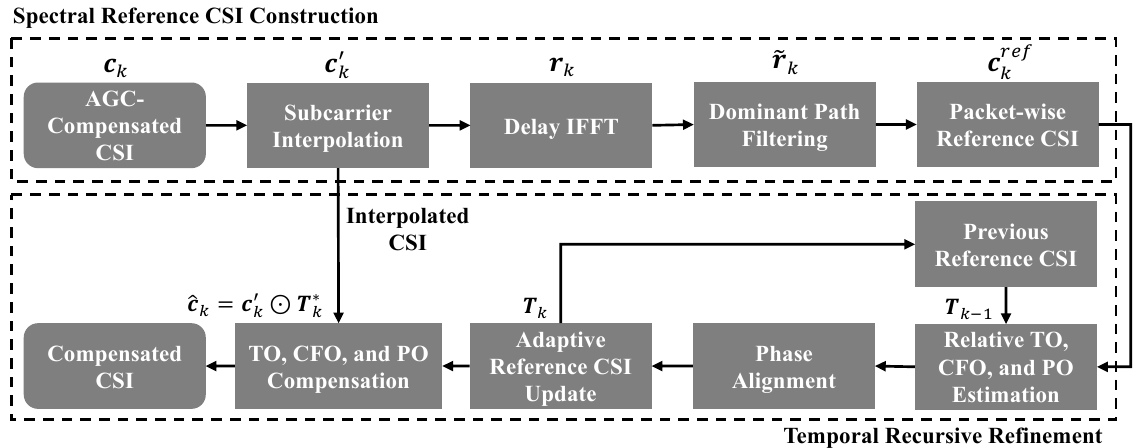}
\caption{Pipeline of the joint spectral-temporal CSI compensation.}
\label{fig:csi_compensation_pipeline}
\vspace{-1.5em}
\end{figure}

To suppress CSI phase distortions while preserving rainfall-sensitive channel perturbations, we propose a CSI compensation framework that jointly exploits the spectral and temporal structures of CSI, as illustrated in Fig.~\ref{fig:csi_compensation_pipeline}.

\subsection{Spectral-Domain Reference CSI Construction}
The key idea is to construct a packet-wise reference from the quasi-static CSI component using subcarrier-domain structure, enabling compensation with a single receive antenna. Since the transmitter and receiver are fixed, the dominant LoS or NLoS component is expected to remain stable over adjacent snapshots and exhibit concentrated delay-domain energy. In contrast, weaker multipath and rainfall-induced perturbations are distributed across delay bins. Isolating the dominant delay component therefore provides a packet-wise reference that retains the TO-, CFO-, and PO-induced phase distortions in the measured CSI, enabling their cancellation in Section~III-B.

\subsubsection{Subcarrier Interpolation}
Let the AGC-compensated CSI at snapshot $k$ be denoted by $\mathbf{c}_k \in \mathbb{C}^{N}$, representing the CSI vector across subcarriers for a single receive antenna. If the subcarrier indices are nonuniform, as in practical systems such as LTE where reference signals are sparsely distributed over frequency, $\mathbf{c}_k$ is interpolated onto a uniform grid, yielding $\mathbf{c}'_k \in \mathbb{C}^{N'}$ for delay-domain processing.

\subsubsection{Delay-Domain Dominant Component Extraction}
The interpolated CSI is transformed into the delay domain using an
$N_{\mathrm{IFFT}}$-point inverse discrete Fourier transform. The resulting
delay-domain response $\mathbf{r}_k\in\mathbb{C}^{N_{\mathrm{IFFT}}}$ has the
$n$-th entry
\begin{equation}
r_{n,k}
=
\frac{1}{N_{\mathrm{IFFT}}}
\sum_{i=1}^{N'}
c'_{i,k}
e^{j2\pi(i-1)\frac{n-1}{N_{\mathrm{IFFT}}}},
\quad
n=1,\ldots,N_{\mathrm{IFFT}}.
\label{eq:delay_transform}
\end{equation}

The dominant delay bin is identified as
\begin{equation}
n^{\max}_k
=
\arg\max_{n}
|r_{n,k}|^2.
\label{eq:main_bin}
\end{equation}
This bin corresponds to the strongest propagation component in the current
snapshot. To emphasize the dominant component while de-emphasizing distant
delay bins, a Gaussian window centered at $n^{\max}_k$ is applied:
\begin{equation}
w_{n,k}
=
\exp\!\left[
-\frac{1}{2}
\left(
\frac{d_{n,k}}{\sigma}
\right)^2
\right],
\label{eq:gaussian_window}
\end{equation}
where
\begin{equation}
d_{n,k}
=
\min\!\left(
|n-n^{\max}_k|,
N_{\mathrm{IFFT}}-|n-n^{\max}_k|
\right)
\label{eq:circular_distance}
\end{equation}
denotes the circular distance from the $n$-th bin to the dominant delay bin.
The parameter $\sigma$ is fixed across snapshots to cover the main lobe and
balance dominant-component retention against multipath suppression, with its
value reported in Section~VI. The filtered delay response is
\begin{equation}
\widetilde{r}_{n,k}
=
w_{n,k}r_{n,k}.
\label{eq:delay_windowed}
\end{equation}

\subsubsection{Packet-Wise Reference CSI Reconstruction}
The packet-wise reference CSI is reconstructed at the $N'$ uniformly spaced subcarrier frequencies using the discrete Fourier transform. The resulting $\mathbf{c}^{\mathrm{ref}}_k\in\mathbb{C}^{N'}$ serves as the packet-wise spectral reference for snapshot $k$, whose $i$-th entry is
\begin{equation}
c^{\mathrm{ref}}_{i,k}
=
\sum_{n=1}^{N_{\mathrm{IFFT}}}
\widetilde{r}_{n,k}
e^{-j2\pi(i-1)(n-1)/N_{\mathrm{IFFT}}},
\quad
i=1,\ldots,N'.
\label{eq:main_path_reconstruct}
\end{equation}
Each snapshot produces its own reference, which is subsequently refined through temporal template alignment and updating. Since delay-domain filtering mainly preserves the dominant coherent component, the reconstructed single-antenna packet-wise reference CSI can be expressed as
\begin{equation}
\mathbf{c}^{\mathrm{ref}}_k
=
e^{-j(\varphi^{\mathrm{CFO}}_k+\phi^{\mathrm{PO}}_k)}
\left(
\mathbf{d}^{\mathrm{TO}}_k
\odot
\mathbf{h}^{\mathrm{ref}}_k
\right),
\label{eq:c_ref_model}
\end{equation}
where the operator $\odot$ is element-wise multiplication,
$\mathbf{d}^{\mathrm{TO}}_k= [e^{-j2\pi f_i\tau^{\mathrm{TO}}_k}]_{i=1}^{N'}
\in\mathbb{C}^{N'}$ is the TO-induced phase vector over the interpolated frequency grid, and $\mathbf{h}^{\mathrm{ref}}_k\in\mathbb{C}^{N'}$ denotes the dominant-component channel vector. The CFO- and PO-induced phases are preserved during delay-domain filtering.

\subsection{Temporal-Domain Reference CSI Refinement}
\label{subsec:temporal_reference_refinement}
For each snapshot, spectral-domain processing extracts the dominant coherent component to form the reference CSI $\mathbf{c}^{\mathrm{ref}}_k$. Limited bandwidth may cause leakage from nearby dynamic components. Because the dominant baseline component evolves smoothly over time, the reference is refined into a snapshot-dependent template $\mathbf{T}_k$ through recursive phase alignment and adaptive updating. This refinement suppresses short-term leakage and stabilizes CSI compensation.

\subsubsection{Template Initialization}
The template is initialized using the first packet-wise spectral reference as
$\mathbf{T}_{1}=\mathbf{c}^{\mathrm{ref}}_{1}$.

\subsubsection{Relative Phase Estimation}
For $k\geq2$, the relative phase distortion between the current reference and
the previous template is obtained from their element-wise conjugate product:
\begin{equation}
\mathbf{q}_{k}
=
\mathbf{c}^{\mathrm{ref}}_{k}
\odot
\mathbf{T}^{*}_{k-1}.
\label{eq:ratio_vector}
\end{equation}
Its $i$-th entry is
$q_{i,k}=c^{\mathrm{ref}}_{i,k}T^{*}_{i,k-1}$,
where $i=1,\ldots,N'$.

Because adjacent snapshots are closely spaced, the dominant reference component
is assumed to remain locally stable, whereas TO, CFO, and PO may vary between
packets. The unwrapped relative phase is therefore approximated as
\begin{equation}
\begin{aligned}
\phi_{i,k}
&=
\left[
\operatorname{unwrap}\!\left(\angle\mathbf{q}_k\right)
\right]_i
\\
&\approx
-2\pi
\xi_i\Delta f
\Delta\tau^{\mathrm{TO}}_k
-
\Delta\varphi^{\mathrm{CFO}}_k
-
\Delta\phi^{\mathrm{PO}}_k
\\
&=
a_k\xi_i+b_k,
\end{aligned}
\label{eq:phase_fit_physical}
\end{equation}
where $\Delta f$ is the subcarrier spacing and $\xi_i$ is the subcarrier index relative to the center frequency, such that $f_i=\xi_i\Delta f$. The operator $\operatorname{unwrap}(\cdot)$ denotes phase unwrapping across subcarriers to remove $2\pi$ discontinuities, while $\Delta\tau^{\mathrm{TO}}_k$, $\Delta\varphi^{\mathrm{CFO}}_k$, and $\Delta\phi^{\mathrm{PO}}_k$ denote the differences between snapshots $k$ and $k-1$. For notational simplicity, the receive-antenna index of $\phi^{\mathrm{PO}}_k$ is omitted throughout the single-antenna derivation.

Accordingly, $a_k$ captures the relative TO-induced phase slope, while $b_k$ contains the aggregate phase offset induced by TO, CFO, and PO. The parameters are estimated by
\begin{equation}
(\widehat{a}_k,\widehat{b}_k)
=
\arg\min_{a,b}
\sum_{i=1}^{N'}
\left(
\phi_{i,k}-a\xi_i-b
\right)^2.
\label{eq:ls_fit}
\end{equation}

\subsubsection{Template Phase Alignment}
The previous template is aligned with the phase frame of the current snapshot as
\begin{equation}
\widetilde{\mathbf{T}}_{k-1}
=
\mathbf{T}_{k-1}
\odot
e^{j(\widehat{a}_k\boldsymbol{\xi}+\widehat{b}_k)}.
\label{eq:template_align}
\end{equation}
where
$\boldsymbol{\xi}=[\xi_1,\ldots,\xi_{N'}]^T$.
This operation removes the relative linear phase trend between adjacent snapshots without separately estimating the TO, CFO, and PO contributions.

\subsubsection{Adaptive Template Update}
The consistency between the current reference and the aligned template is measured by the normalized complex correlation
\begin{equation}
\varrho_k
=
\frac{
\left|
(\mathbf{c}^{\mathrm{ref}}_k)^H
\widetilde{\mathbf{T}}_{k-1}
\right|
}{
\|\mathbf{c}^{\mathrm{ref}}_k\|_2
\|\widetilde{\mathbf{T}}_{k-1}\|_2
}.
\label{eq:cosine_similarity}
\end{equation}
The adaptive update coefficient is
\begin{equation}
\beta_k
=
\beta_{\min}
+
(\beta_{\max}-\beta_{\min})
\varrho_k,
\label{eq:adaptive_beta}
\end{equation}
where
$0<\beta_{\min}<\beta_{\max}<1$
are preset bounds. The refined template is updated as
\begin{equation}
\mathbf{T}_k
=
\beta_k\widetilde{\mathbf{T}}_{k-1}
+
(1-\beta_k)\mathbf{c}^{\mathrm{ref}}_k.
\label{eq:template_update}
\end{equation}
A larger $\beta_k$ provides stronger temporal smoothing when the references are consistent, whereas a smaller value enables faster adaptation. The bounds on $\beta_k$ balance the contributions of the aligned template and the current reference.

\subsubsection{TO, CFO, and PO Compensation}
The compensated CSI is obtained through template conjugate multiplication:
\begin{equation}
\widehat{\mathbf{c}}_k
=
\mathbf{c}'_k
\odot
\mathbf{T}^{*}_k.
\label{eq:csi_compensated}
\end{equation}

For the single-antenna derivation, let
$\mathbf{h}^{\mathrm{B}}_k$, $\mathbf{h}^{\mathrm{X}}_k$, and $\mathbf{h}^{\mathrm{R}}_k\in\mathbb{C}^{N'}$ denote the interpolated subcarrier-domain channel vectors corresponding to one receive antenna, extracted from $\mathbf{H}^{\mathrm{B}}_k$, $\mathbf{H}^{\mathrm{X}}_k$, and $\mathbf{H}^{\mathrm{R}}_k$, respectively. The interpolated CSI is expressed as
\begin{equation}
\mathbf{c}'_k
=
e^{-j(\varphi^{\mathrm{CFO}}_k+\phi^{\mathrm{PO}}_k)}
\left[
\mathbf{d}^{\mathrm{TO}}_k
\odot
\left(
\mathbf{h}^{\mathrm{B}}_k
+
\mathbf{h}^{\mathrm{X}}_k
+
\mathbf{h}^{\mathrm{R}}_k
\right)
\right].
\label{eq:csi_single_antenna_model}
\end{equation}
The refined template is approximated as
\begin{equation}
\mathbf{T}_k
\approx
e^{-j(\varphi^{\mathrm{CFO}}_k+\phi^{\mathrm{PO}}_k)}
\left(
\mathbf{d}^{\mathrm{TO}}_k
\odot
\widehat{\mathbf{h}}^{\mathrm{ref}}_k
\right),
\label{eq:template_model}
\end{equation}
where $\widehat{\mathbf{h}}^{\mathrm{ref}}_k$ is the temporally refined dominant reference component. Substituting Eqs.~\eqref{eq:csi_single_antenna_model} and \eqref{eq:template_model} into Eq.~\eqref{eq:csi_compensated} yields
\begin{equation}
\widehat{\mathbf{c}}_k
\approx
\left(
\mathbf{h}^{\mathrm{B}}_k
+
\mathbf{h}^{\mathrm{X}}_k
+
\mathbf{h}^{\mathrm{R}}_k
\right)
\odot
\left(
\widehat{\mathbf{h}}^{\mathrm{ref}}_k
\right)^*.
\label{eq:csi_compensated_expand}
\end{equation}
Thus, conjugate multiplication cancels the TO-, CFO-, and PO-induced phase distortions. Because the template is derived from the dominant path, rainfall-induced phase perturbations embedded in that path are also suppressed, while its magnitude variation is not explicitly normalized out. The compensated CSI therefore emphasizes multipath and rainfall-induced dynamics relative to the dominant baseline component and is used for delay, Doppler, and angle-domain feature extraction.

\section{Physics-Guided Rainfall Feature Extraction}
After CSI compensation, rainfall-induced variations remain embedded in propagation and environmental components. The compensated CSI is transformed into delay--Doppler representations, with optional angle-domain extension for multi-antenna receivers, to suppress dynamic interference.

\subsection{Per-Subcarrier Temporal Normalization}
Frequency-selective fading causes unequal CSI amplitudes across subcarriers, which may mask rainfall perturbations on low-energy subcarriers. For a CPI containing $L$ compensated CSI snapshots, we form
\begin{equation}
\widehat{\mathbf{X}}
=
\left[
\widehat{\mathbf{c}}_{1},
\widehat{\mathbf{c}}_{2},
\ldots,
\widehat{\mathbf{c}}_{L}
\right]
\in
\mathbb{C}^{N'\times L}.
\label{eq:compensated_csi_matrix}
\end{equation}
The subcarrier-wise temporal mean is
\begin{equation}
\bar{\mathbf{x}}
=
\frac{1}{L}
\widehat{\mathbf{X}}\mathbf{1},
\label{eq:csi_mean}
\end{equation}
where $\mathbf{1}\in\mathbb{R}^{L}$ is an all-one vector and $\bar{\mathbf{x}}\in\mathbb{C}^{N'}$ estimates the baseline response. The normalized CSI is denoted as
\begin{equation}
\widetilde{\mathbf{X}}
=
\left(
\widehat{\mathbf{X}}
-
\bar{\mathbf{x}}\mathbf{1}^{T}
\right)
\oslash
|\bar{\mathbf{x}}|\mathbf{1}^{T},
\label{eq:mean_remove_csi}
\end{equation}
where $\oslash$ denotes element-wise division. This operation removes the local baseline and balances subcarrier amplitudes without introducing additional phase rotation.

\subsection{Beamforming-Based Delay--Doppler Feature Extraction}

\subsubsection{Delay Focusing}
Beamforming \cite{van2002optimum} is first applied across the  subcarriers. It preserves the selected delay-bin response while suppressing leakage and structured interference from other delay components. The covariance matrix is formed as $\mathbf{R}_{xx}=\widetilde{\mathbf{X}}\widetilde{\mathbf{X}}^{H}$. Forward--backward averaging and diagonal loading are applied to improve numerical stability:
\begin{equation}
\widetilde{\mathbf{R}}_{xx}
=
\frac{1}{2}
\left(
\mathbf{R}_{xx}
+
\mathbf{J}\mathbf{R}_{xx}^{*}\mathbf{J}
\right)
+
\delta_{\mathrm{DL}}\mathbf{I},
\label{eq:Rxx_loaded}
\end{equation}
where $\mathbf{J}$ is the reversal matrix, $\mathbf{I}$ is the identity matrix, and $\delta_{\mathrm{DL}}>0$ is the diagonal loading parameter.

Let $\{\tau_r\}_{r=1}^{N_{\tau}}$ denote the candidate delay grid. The steering vector for the $r$-th delay bin is
\begin{equation}
\mathbf{b}_r
=
\left[
e^{-j2\pi f_1\tau_r},
e^{-j2\pi f_2\tau_r},
\ldots,
e^{-j2\pi f_{N'}\tau_r}
\right]^T,
\label{eq:range_steering}
\end{equation}
The beamforming weight is
\begin{equation}
\mathbf{w}_r
=
\frac{
\widetilde{\mathbf{R}}_{xx}^{-1}\mathbf{b}_r
}{
\mathbf{b}_r^{H}
\widetilde{\mathbf{R}}_{xx}^{-1}\mathbf{b}_r
}.
\label{eq:mvdr_weight}
\end{equation}

Applying the beamforming weight to the $k$-th normalized CSI snapshot gives
\begin{equation}
z_{r,k}
=
\mathbf{w}_{r}^{H}\widetilde{\mathbf{x}}_k,
\label{eq:focused_signal}
\end{equation}
where $\widetilde{\mathbf{x}}_k\in\mathbb{C}^{N'}$ is the $k$-th column of $\widetilde{\mathbf{X}}$.

\subsubsection{Doppler Feature Extraction}
The slow-time sequence for the $r$-th delay bin is
\begin{equation}
\mathbf{z}_r
=
[z_{r,1},z_{r,2},\ldots,z_{r,L}]^T
\in\mathbb{C}^{L}.
\label{eq:slow_time_sequence}
\end{equation}
Let $\{f_{\mathrm{D},m}\}_{m=1}^{N_{\mathrm{D}}}$ denote the Doppler frequency grid, where $N_{\mathrm{D}}$ is the number of Doppler bins. The Doppler response is obtained using an $N_{\mathrm{D}}$-point slow-time FFT:
\begin{equation}
p_{r,m}
=
\sum_{k=1}^{L}
z_{r,k}
e^{-j2\pi f_{\mathrm{D},m}(k-1)\Delta t}.
\label{eq:doppler_fft}
\end{equation}
Its magnitude forms the delay--Doppler feature:
\begin{equation}
D_{r,m}
=
|p_{r,m}|,
\quad
r=1,\ldots,N_{\tau},\;
m=1,\ldots,N_{\mathrm{D}}.
\label{eq:doppler_feature}
\end{equation}
Stacking all delay and Doppler bins yields
\begin{equation}
\mathbf{D}^{\mathrm{DD}}
=
[D_{r,m}]
\in
\mathbb{R}_{+}^{N_{\tau}\times N_{\mathrm{D}}},
\label{eq:delay_doppler_map}
\end{equation}
which characterizes the delay--Doppler distribution of channel fluctuations within one CPI.

\subsection{Optional Extension to Multi-Antenna Receivers}
Although developed for single-antenna receivers, the proposed framework can be extended to an $M$-antenna array. The delay--Doppler processing is first applied independently to each receive-antenna CSI stream. For each delay--Doppler bin $(r,m)$, the resulting antenna responses are stacked as
\begin{equation}
\mathbf{p}_{r,m}
=
\left[
p^{(1)}_{r,m},
p^{(2)}_{r,m},
\ldots,
p^{(M)}_{r,m}
\right]^T.
\label{eq:prm_stack}
\end{equation}

For a uniform linear array, an antenna-domain FFT is then applied to extract the angle response:
\begin{equation}
a_{q,r,m}
=
\sum_{a=1}^{M}
p^{(a)}_{r,m}
e^{-j2\pi(a-1)u_q},
\label{eq:aoa_fft}
\end{equation}
where $u_q$ denotes the spatial frequency associated with the $q$-th angle bin. The delay--Doppler--AoA feature cube is
\begin{equation}
\mathbf{D}^{\mathrm{DDA}}
=
\left[
|a_{q,r,m}|
\right]
\in
\mathbb{R}_{+}^{N_{\theta}\times N_{\tau}\times N_{\mathrm{D}}},
\label{eq:delay_doppler_aoa_cube}
\end{equation}
which provides additional spatial discrimination when multiple receive antennas are available.

\section{Rainfall Classification and Estimation}
The extracted delay--Doppler features contain both rainfall-sensitive responses and environmental interference. Practical rainfall sensing therefore applies range and Doppler filtering followed by baseline normalization to suppress interference and setup-dependent variations. Lightweight supervised models are used for rainfall classification and intensity estimation.

\subsection{Physics-Guided Feature Filtering}
\subsubsection{Delay}
Delay reflects propagation distance and determines the effective signal interaction length with rain. Short-delay components are generally less sensitive because the signal traverses a smaller rain volume. Longer-delay components may exhibit stronger rainfall-induced perturbations, but are often weaker and more susceptible to noise, multipath, and dynamic interference. Therefore, delay provides coarse propagation discrimination for selecting rainfall-sensitive paths.

\subsubsection{Doppler}
Rainfall produces weak but persistent dynamic scattering in the Doppler domain. Compared with delay, whose resolution is limited by communication bandwidth, Doppler processing exploits slow-time accumulation within one CPI to achieve finer resolution. Environmental interference often exhibits distinct velocity signatures, allowing interference-dominant regions to be suppressed through Doppler filtering.

Based on these physical insights, the extracted features are filtered over delay and Doppler to enhance rainfall-sensitive responses and suppress environmental interference. The selected delay-bin set is defined as
\begin{equation}
\mathcal{D}
=
\left\{
r:\ \tau_{\min}\le \tau_r\le \tau_{\max}
\right\},
\label{eq:delay_gate}
\end{equation}
where $\tau_r$ denotes the delay associated with the $r$-th bin. This selection retains delay bins with reliable rainfall sensitivity while suppressing unreliable short- and long-delay components. The selected Doppler-bin set is defined as
\begin{equation}
\mathcal{F}_{\mathrm{D}}
=
\left\{
m:\ v_{\min}\le |v_m|\le v_{\max}
\right\},
\label{eq:doppler_gate}
\end{equation}
where $v_m$ is the Doppler-equivalent velocity of the $m$-th Doppler bin. This filtering suppresses interference-dominant Doppler regions while retaining rainfall-sensitive responses.

Since Doppler captures persistent rainfall-induced temporal fluctuations and provides finer discrimination than delay under limited communication bandwidth, the filtered Doppler profile is used as the primary rainfall-sensitive feature. For the $\kappa$-th CPI, it is obtained by averaging over the selected delay bins:
\begin{equation}
P_{m,\kappa}
=
\frac{1}{|\mathcal{D}|}
\sum_{r\in\mathcal{D}}
D_{r,m}^{(\kappa)},
\qquad
m\in\mathcal{F}_{\mathrm{D}},
\label{eq:selected_doppler_profile}
\end{equation}
where $\kappa$ denotes the CPI index. The resulting one-dimensional Doppler profile is denoted by $\mathbf{P}_{\kappa}=[P_{m,\kappa}]_{m\in\mathcal{F}_{\mathrm{D}}}$.

When a multi-antenna receiver is available, AoA information can provide additional spatial discrimination against environmental interference. Angle-domain filtering can retain rainfall-sensitive directions while suppressing regions dominated by localized motion. However, its resolution is often limited by the number of receive antennas and the array aperture. AoA filtering is therefore treated as an optional extension and is not used in the reported WiFi and LTE rainfall metrics; its quantitative evaluation is left for future work.

\subsection{Doppler Feature Generalization Across Deployment Setups}
Raw Doppler responses are not directly comparable across deployment setups because their magnitude and temporal baseline depend on both link configuration and background dynamics. Differences in transceiver placement, propagation distance, viewing geometry, and surrounding scatterers can alter the Doppler magnitude even under similar rainfall conditions, limiting the transferability across sites or links. Meanwhile, background motion, traffic-load changes, and evolving multipath can cause the non-rain Doppler response to drift over time. In stable environments, this drift is slow enough to support normalization against a long-term baseline. In time-varying environments, however, such a baseline may become unreliable, motivating normalization based on the Doppler distribution within each CPI. Accordingly, two complementary strategies are adopted: baseline-normalized Doppler features for stable backgrounds and CPI-wise Doppler-shape features for time-varying backgrounds. Both are computed online from CPI-level Doppler profiles without storing raw CSI, making them suitable for real-time rainfall sensing.

\subsubsection{Baseline-Normalized Doppler Feature for Stable Backgrounds}

For stable environments, the Doppler magnitude is normalized relative to a slowly varying historical baseline. The one-dimensional Doppler profile of CPI $\kappa$ is first averaged over the selected rainfall-sensitive Doppler bins:
\begin{equation}
P_{\kappa}
=
\frac{1}{|\mathcal{F}_{\mathrm{D}}|}
\sum_{m\in\mathcal{F}_{\mathrm{D}}}
P_{m,\kappa},
\label{eq:mean_doppler_power}
\end{equation}
where $P_{m,\kappa}$ is the Doppler magnitude at the $m$-th bin of CPI $\kappa$. Let $\sigma_{\kappa}^{\mathrm{P}}$ denote the standard deviation of $\{P_{m,\kappa}\}_{m\in\mathcal{F}_{\mathrm{D}}}$.

Before processing CPI $\kappa$, the historical baseline is represented by $\mu_{\kappa-1}$ and $\sigma_{\kappa-1}$. The baseline-normalized deviation and its squared-deviation indicator are computed as
\begin{equation}
z_{\kappa}
=
\frac{
P_{\kappa}-\mu_{\kappa-1}
}{
\sigma_{\kappa-1}
},
\qquad
\chi_{\kappa}^2
=
z_{\kappa}^2.
\label{eq:chi_square_indicator}
\end{equation}
This normalization expresses the current Doppler magnitude as a relative deviation from the historical background.

After computing the current indicators, the long-term baseline mean and variability are recursively updated as
\begin{equation}
\mu_{\kappa}
=
(1-\eta)\mu_{\kappa-1}
+
\eta P_{\kappa},
\quad
\sigma_{\kappa}
=
(1-\eta)\sigma_{\kappa-1}
+
\eta \sigma_{\kappa}^{\mathrm{P}},
\label{eq:baseline_statistics}
\end{equation}
where $0<\eta\ll1$ is the update rate. A small $\eta$ tracks gradual baseline drift while limiting the absorption of short-term rainfall-induced fluctuations. The initial values $\mu_0$ and $\sigma_0$ are estimated from an initial CPI.

Even in stable environments, environmental changes, hardware drift, and long-term multipath variations may cause gradual baseline variation. To suppress such slow changes and emphasize short-term rainfall-induced dynamics, the squared difference between adjacent indicators is computed as
\begin{equation}
\Delta\chi_{\kappa}^2
=
\left(
\chi_{\kappa}^2-\chi_{\kappa-1}^2
\right)^2.
\label{eq:chi_square_diff}
\end{equation}
During no-rain periods, $\Delta\chi_{\kappa}^2$ generally remains small and exhibits only weak temporal variation, whereas rainfall-induced Doppler fluctuations produce larger adjacent variations.

\subsubsection{Doppler-Shape Normalization for Time-Varying Backgrounds}
For time-varying environments, a long-term no-rain baseline may become unreliable because environmental motion, scheduling variations, and evolving multipath can continuously alter the Doppler magnitude. The Doppler response is therefore characterized using statistics computed independently within each CPI. Let
\begin{equation}
\mathbf{P}_{\kappa}
=
\left[
P_{m,\kappa}
\right]_{m\in\mathcal{F}_{\mathrm{D}}}
\label{eq:selected_doppler_vector}
\end{equation}
denote the selected Doppler profile of CPI $\kappa$, and let $\mu_{\kappa}^{\mathrm{P}}$ and $\sigma_{\kappa}^{\mathrm{P}}$ denote its mean and standard deviation, respectively. The coefficient of variation (CV) is defined as
\begin{equation}
\mathrm{CV}_{\kappa}
=
\frac{
\sigma_{\kappa}^{\mathrm{P}}
}{
\mu_{\kappa}^{\mathrm{P}}
},
\label{eq:doppler_cv}
\end{equation}
Unlike the baseline-normalized indicator, $\mathrm{CV}_{\kappa}$ depends only on the CPI and measures the relative dispersion of the Doppler profile rather than its absolute magnitude. It is less sensitive to deployment-dependent amplitude scaling caused by transceiver placement, link geometry, and antenna gain. To capture the temporal evolution of the Doppler shape across consecutive CPIs, the squared change in the CV is defined as
\begin{equation}
\Delta\mathrm{CV}_{\kappa}^2
=
\left(
\mathrm{CV}_{\kappa}
-
\mathrm{CV}_{\kappa-1}
\right)^2.
\label{eq:doppler_cv_diff_square}
\end{equation}
When adjacent Doppler profiles have similar shapes, $\Delta\mathrm{CV}_{\kappa}^2$ remains small over consecutive time intervals. Rainfall-induced scattering can perturb the Doppler distribution across adjacent CPIs, resulting in larger short-term shape variations.

While $\mathrm{CV}_{\kappa}$ and $\Delta\mathrm{CV}_{\kappa}^2$ provide scalar summaries of the Doppler shape and its temporal variation, the normalized Doppler profile is retained to preserve detailed bin-wise information. It is computed within each CPI as
\begin{equation}
\widetilde{P}_{m,\kappa}
=
\frac{
P_{m,\kappa}-\mu_{\kappa}^{\mathrm{P}}
}{
\sigma_{\kappa}^{\mathrm{P}}
},
\qquad
m\in\mathcal{F}_{\mathrm{D}},
\label{eq:normalized_doppler_shape}
\end{equation}
and the resulting profile is
\begin{equation}
\widetilde{\mathbf{P}}_{\kappa}
=
\left[
\widetilde{P}_{m,\kappa}
\right]_{m\in\mathcal{F}_{\mathrm{D}}}.
\label{eq:normalized_doppler_profile_vector}
\end{equation}
The profile $\widetilde{\mathbf{P}}_{\kappa}$ preserves Doppler-shape information, including spectral broadening, asymmetry, and local fluctuations, while reducing dependence on absolute signal magnitude. Therefore, $\mathrm{CV}_{\kappa}$, $\Delta\mathrm{CV}_{\kappa}^2$, and $\widetilde{\mathbf{P}}_{\kappa}$ characterize the Doppler shape and its temporal evolution in time-varying environments.

\subsubsection{Comparison of the Two Normalization Strategies}
The two strategies provide different tradeoffs between rainfall sensitivity and deployment robustness. In stable environments, baseline normalization preserves absolute Doppler deviations and their temporal variations, providing stronger rainfall discrimination when a reliable no-rain reference is available. In time-varying environments, CPI-wise Doppler-shape normalization suppresses background drift and deployment-dependent amplitude scaling, improving robustness at the cost of discarding part of the absolute-magnitude information. The performance tradeoff is evaluated in Section~\ref{sec:experimental_results}.

\begin{figure*}[t]
    \centering
    \begin{subfigure}[t]{0.69\linewidth}
        \centering
        \includegraphics[width=\linewidth]{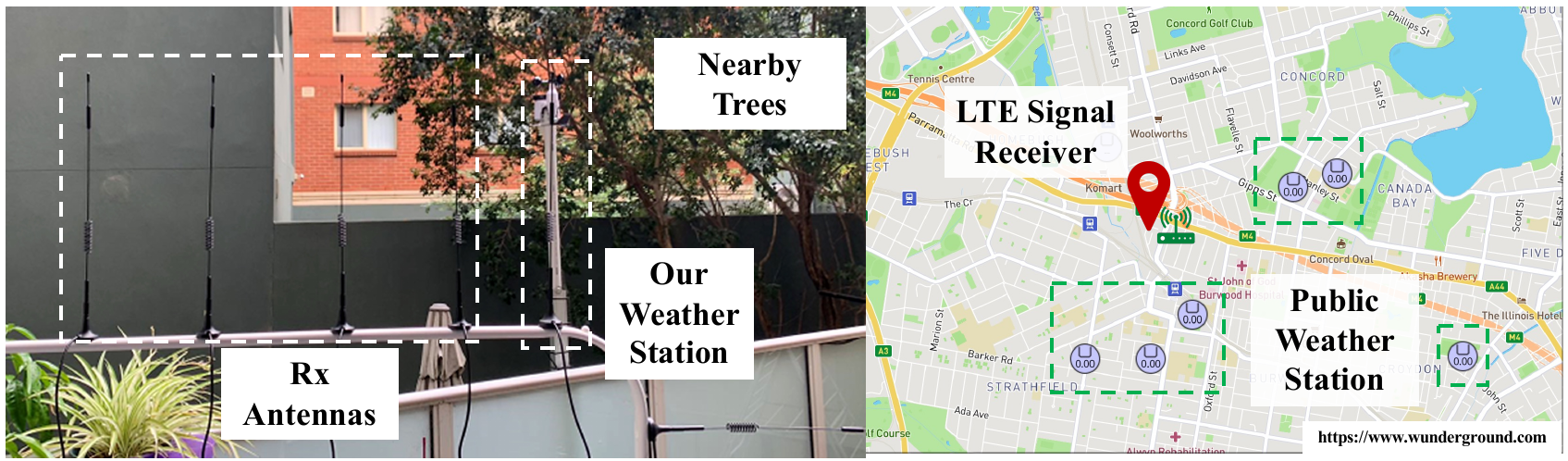}
        \caption{Sub-6~GHz LTE rainfall-monitoring setup.}
        \label{fig:lte_setup}
    \end{subfigure}
    \hfill
    \begin{subfigure}[t]{0.3\linewidth}
        \centering
        \includegraphics[width=\linewidth]{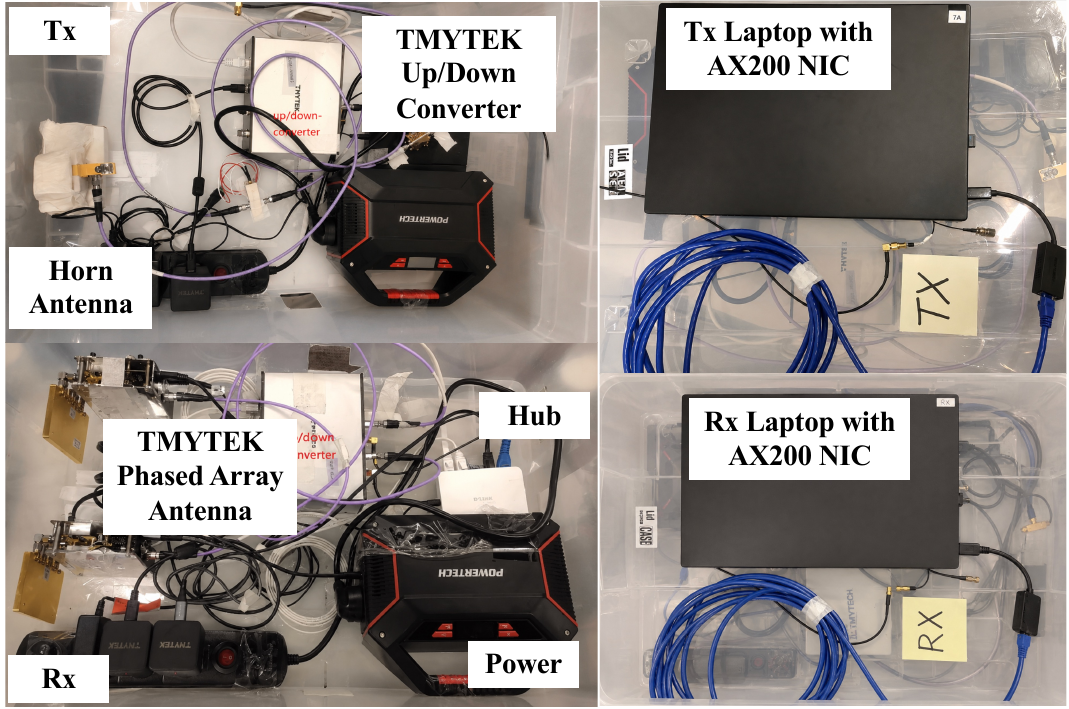}
        \caption{28 GHz mmWave platform.}
        \label{fig:mmwave_setup}
    \end{subfigure}
    \caption{Experimental platforms for sub-6~GHz LTE monitoring and 28~GHz mmWave rainfall measurements.}
    \label{fig:experimental_setups}
    \vspace{-1.5em}
\end{figure*}

\subsection{Rainfall Classification and Intensity Estimation}
After Doppler feature filtering and normalization, the framework constructs compact rainfall-sensitive features according to background stability. For stable environments, baseline-normalized Doppler indicators are used for rainfall-level classification. For time-varying environments, Doppler-shape features and temporal variation indicators are used for rainfall-intensity estimation. In both cases, lightweight supervised models are adopted to learn the mapping from the extracted physics-guided features to the corresponding rainfall outputs.

\subsubsection{Rainfall-Level Classification under Stable Backgrounds}
This task targets deployments with limited ground-level interference, such as high-mounted links between infrastructure nodes, where vehicle, pedestrian, and nearby-object motion has only a limited effect on the measured channel. In such deployments, rainfall produces Doppler fluctuations relative to a slowly varying environmental baseline. Compared with $\Delta\mathrm{CV}_{\kappa}^{2}$, $\Delta\chi_{\kappa}^{2}$ exhibits greater sensitivity to rainfall, as demonstrated in Section~\ref{sec:wifi_classification_results}. Rainfall-level classification then uses statistical descriptors of $\Delta\chi_{\kappa}^{2}$ over a sliding window of $W$ consecutive CPIs. The feature vector for CPI $\kappa$ is defined as
\begin{equation}
\resizebox{\linewidth}{!}{$
\displaystyle
\mathbf{u}_{\kappa}
=
\Phi
\left(
\left[
\Delta\chi_{\kappa-W+1}^{2},
\ldots,
\Delta\chi_{\kappa}^{2}
\right]
\right)
=
\left[
u_{1,\kappa},
u_{2,\kappa},
\ldots,
u_{N_u,\kappa}
\right],
$}
\label{eq:classification_feature_vector}
\end{equation}
where $W$ denotes the number of consecutive CPIs included in the sliding window, $N_u$ is the number of descriptors, and $\Phi(\cdot)$ extracts conventional statistics, including the mean, standard deviation, median, maximum, and percentile values. The rainfall level is classified as
\begin{equation}
\widehat{y}_{\kappa}
=
f_{\mathrm{cls}}
\left(
\mathbf{u}_{\kappa}
\right),
\label{eq:rainfall_classifier}
\end{equation}
where $\widehat{y}_{\kappa}\in\{0,1,2\}$ represents no rain, light rain, and heavy rain, respectively. In this work, a Random Forest classifier \cite{scornet2026theory} is adopted to learn the nonlinear mapping from the compact statistical descriptors to the rainfall level.

\subsubsection{Rainfall-Intensity Estimation under Time-Varying Backgrounds}

For rainfall-intensity estimation under time-varying backgrounds, the normalized one-dimensional Doppler profile is directly used as the channel input. Unlike scalar indicators that summarize only selected Doppler characteristics, the full profile preserves the distribution of channel fluctuations across Doppler bins. A selected Doppler range is retained to suppress background fluctuations and interference unrelated to rainfall. Each profile is independently normalized to reduce link-dependent magnitude variations while retaining its relative Doppler-domain structure. To capture the accumulation and short-term evolution of rainfall effects, the estimator jointly processes the profiles from the most recent $L_{\mathrm{est}}$ sensing intervals. The rainfall intensity at CPI $\kappa$ is estimated as
\begin{equation}
\widehat{r}_{\kappa}
=
f_{\mathrm{est}}
\left(
\left\{
\widetilde{\mathbf{P}}_{i}
\right\}_{i=\kappa-L_{\mathrm{est}}+1}^{\kappa}
\right),
\label{eq:rainfall_estimation}
\end{equation}
where $L_{\mathrm{est}}$ is the number of consecutive sensing intervals, $\widetilde{\mathbf{P}}_{i}$ denotes the normalized Doppler profile from the selected Doppler bins at CPI $i$, $\widehat{r}_{\kappa}\in\mathbb{R}_{+}$ is the estimated rainfall intensity, and $f_{\mathrm{est}}(\cdot)$ denotes a lightweight one-dimensional convolutional estimator. By processing consecutive profiles, the estimator captures the Doppler-bin distribution within each sensing interval and its temporal evolution across intervals. 

Additionally, RSRP can be processed with the Doppler profiles as an auxiliary input, as evaluated in the subsequent comparison. The derived indicators $\Delta\chi_{\kappa}^{2}$ and $\Delta\mathrm{CV}_{\kappa}^{2}$ are not used as estimator inputs, but provide interpretable summaries of rainfall-related variations contained in the Doppler profiles. They characterize temporal changes in baseline-relative Doppler power and profile dispersion, respectively. Their relationships with rainfall intensity are examined in Section~\ref{sec:lte_rainfall_estimation_results}, providing physical interpretation for the use of the complete Doppler profile in rainfall-intensity estimation.

\section{Implementation}
\label{sec:implementation}

\subsection{Rainfall Datasets}
\subsubsection{WiFi Rainfall Dataset}
An outdoor WiFi CSI dataset is used for three-class rainfall classification. The measurements are conducted in a residential-yard environment containing vegetation, building reflections, and occasional human activity, while ground-level dynamic interference remains limited during most measurement periods. CSI is collected using a 5~GHz Intel 5300 WiFi platform with three receive antennas. The receiver operates in monitor mode using the Linux 802.11n CSI Tool \cite{halperin2011tool}. Measurements are collected from September~8,~2023 to January~15,~2024 under non-rainy and rainy conditions, with the transmitter and receiver placed at different locations covering both LoS and NLoS links. A tipping-bucket rain gauge with a resolution of 0.2794~mm per tip provides the rainfall reference. Each measurement lasts 10--15 minutes, and the total dataset size is approximately 10.7~GB. Since the rain gauge records cumulative rainfall depth and updates only when a bucket tip occurs, its readings are quantized and may remain unchanged over short intervals, particularly during light rainfall. To align the rainfall reference with each CSI sensing interval, the rain rate is estimated by dividing the change between consecutive cumulative readings by the elapsed time, yielding an interval-averaged rain rate. Zero values are labeled as \textit{No Rain}, while positive rain rates are divided at their training-set median into \textit{Light Rain} and \textit{Heavy Rain}, with the same threshold applied to the testing data.

\subsubsection{Real-World LTE Rainfall Dataset}
An outdoor LTE CSI dataset is collected from commercial LTE downlink signals using a software-defined radio (SDR)-based receiver. As shown in Fig.~\ref{fig:lte_setup}, the receiver is deployed near a weather station in a near-ground urban environment and uses two receive antennas. The receiver gain is fixed during data collection to avoid amplitude variations caused by AGC. For each LTE cell, CSI is estimated from two cell-specific reference signal (CRS) transmit antenna ports, denoted by TX0 and TX1, and two receive antennas, yielding two receive streams per transmit port. The receive antennas are not calibrated as an array; their spacing and relative phase are not used, and no angle-domain processing is performed. Each receive stream is processed independently for CSI compensation and Doppler-feature extraction. The measured carrier frequencies are 0.763, 0.778, 0.875, 1.8125, 1.840, 1.8575, 2.1175, 2.1399, 2.165, 2.6598, and 2.68~GHz. Continuous measurements were conducted in Homebush, Sydney, from June~30 to July~10,~2024. Each CSI snapshot contains 12 complex channel estimates distributed across a bandwidth of 1.08~MHz, with a frequency spacing of 90~kHz between adjacent estimates. Under the normal cyclic-prefix configuration, each 0.5-ms LTE slot contains seven OFDM symbols, two of which carry CRS for each transmit port, thereby forming two temporally interleaved CSI sequences. Each sequence repeats every 0.5~ms, resulting in an average combined snapshot interval of approximately 0.25~ms. Each measurement file contains approximately 3,960 CSI snapshots and covers about 1 s, while consecutive files are recorded at approximately 3-minute intervals. 

\subsubsection{mmWave Rainfall Dataset}
A 28~GHz mmWave CSI dataset is collected to examine rainfall-associated channel responses at high carrier frequencies. As shown in Fig.~\ref{fig:mmwave_setup}, the system combines Intel AX200 WiFi NICs with TMYTEK mmWave up/down-conversion and phased-array modules. A horn antenna is used at the transmitter, while two phased-array modules are used at the receiver, forming a 1Tx--2Rx bistatic configuration. Measurements are collected over a bandwidth of 160~MHz. Owing to the stronger interaction between millimeter-wave signals and raindrops, this dataset provides clearer observations of CSI variations and is used to verify whether the extracted features capture channel dynamics.

\subsection{CSI Compensation and Feature Extraction}
\subsubsection{WiFi Setting}
For WiFi, the 30 measured CSI subcarriers are interpolated onto 57 uniformly spaced subcarriers indexed from $-28$ to $28$. A 128-point inverse Fourier transform and a Gaussian window with $\sigma=64$ are used to construct the dominant-path reference, with $\beta_{\min}=0.6$ and $\beta_{\max}=0.9$ for temporal refinement. The baseline update rate is set to $\eta=0.01$. CSI snapshots are grouped into 128-packet CPIs at 1~kHz. Delay focusing uses 64 bins corresponding to excess path lengths within $[0,32]$~m, followed by 128-bin Doppler processing. Doppler components with velocity magnitudes between 5 and 25~m/s, i.e., $5\leq |v|\leq25$~m/s, are retained.

\subsubsection{LTE Setting}
For LTE, a 32-point inverse Fourier transform and a Gaussian window with $\sigma=16$ are used for dominant-path extraction, followed by the same temporal refinement and baseline-update settings as in the WiFi case. Capon delay focusing uses 32 bins corresponding to excess path lengths within $[0,128]$~m, and Doppler processing employs a 2048-point zoom FFT. Based on empirical observations of the measured background, Doppler components satisfying $20\leq |v|\leq200$~m/s are retained to suppress near-zero background fluctuations and high-velocity interference.

\subsubsection{mmWave Setting}
For mmWave, CSI is processed at 28~GHz using the same temporal refinement, CPI construction, and delay-focusing settings as in the WiFi case. A 256-point inverse Fourier transform is used for dominant-path extraction, followed by a 128-bin zoom FFT over the Doppler-velocity interval $|v|\leq2$~m/s. Processing is performed independently for each receive antenna, after which the delay--Doppler magnitudes are averaged over the delay bins to obtain the Doppler--time representation for qualitative rainfall analysis.

\subsection{Rainfall Classification and Intensity Estimation}
\subsubsection{WiFi Rainfall Classification}
Classification samples are constructed using a 500-CPI sliding window with a step size of one CPI, such that each sample summarizes approximately 64 seconds of observations. For each window, the mean, standard deviation, median, maximum, and 90th percentile of $\Delta\chi_{\kappa}^{2}$ are extracted, yielding a five-dimensional feature vector. These features characterize the short-term temporal variation of the baseline-normalized Doppler response. A Random Forest classifier is adopted as the main model, and class-balanced sample weights are applied during training. File-level group-holdout validation is used, with 30\% of the measurement files reserved for testing, and all overlapping windows from the same file are assigned to the same split to prevent temporal leakage.

\subsubsection{LTE Rainfall Intensity Estimation}
A lightweight neural estimator is used for LTE rainfall-intensity estimation. Each input sample combines selected one-dimensional Doppler profiles from three consecutive sensing intervals to capture the short-term evolution of rainfall-induced channel fluctuations. The rainfall labels are temporally averaged over five intervals, corresponding to approximately 15 minutes, to reduce the quantization and timing uncertainty of the tipping-bucket measurements and improve alignment with the accumulated channel response. The profile sequence is processed by one-dimensional convolutional layers and fully connected layers to estimate rainfall intensity. The estimator is trained for 300 epochs using AdamW with an initial learning rate of $1\times10^{-3}$ and cosine-annealing scheduling. A weighted MSE loss is employed to reduce the influence of the imbalanced rainfall distribution. All transmit-port links except the 2.68~GHz TX1 link are used for training, while this link is reserved for testing. Since TX0 and TX1 correspond to different CRS antenna ports and provide independently estimated port-specific CSI, the test data represent a held-out effective transmit link rather than duplicated observations of a training link. After feature extraction and three-interval sample construction, the dataset contains 98,672 training samples and 4,698 testing samples.

\subsection{Baseline Comparisons and Ablation Studies}
\subsubsection{WiFi Classification}
For WiFi rainfall classification, the temporal variations of the RSSI and CSI amplitude are compared with $\Delta\chi_{\kappa}^{2}$ to examine their sensitivity to rainfall-induced channel fluctuations. Random Forest is adopted as the main classifier, while Decision Tree and gradient-boosting classifiers are included to evaluate the influence of the learning model under the same data split. Classification performance is evaluated using precision, recall, and F1-score.

\subsubsection{LTE Estimation}
For LTE rainfall-intensity estimation, RSRP is used as the conventional radio-measurement baseline because it characterizes the received power of LTE reference signals. The ablation study compares the selected Doppler profile, the full Doppler profile, their combinations with RSRP, and the RSRP-only input under the same training and testing split. Performance is evaluated using the mean absolute error (MAE). These comparisons evaluate the benefit of Doppler-range selection and demonstrate that the Doppler profiles provide more robust rainfall-intensity estimation than RSRP. Seasonal-trend decomposition using LOESS (STL) \cite{cleveland1990stl} is additionally applied to the logarithmically transformed $\Delta\chi_{\kappa}^{2}$ and the $\Delta\mathrm{CV}_{\kappa}^{2}$ sequences for temporal interpretation. By separating each sequence into trend, daily seasonal, and residual components, STL attenuates recurring variations caused by traffic, pedestrian activity, vegetation motion, transmission scheduling, and slowly varying multipath conditions, making short-term rainfall-related variations easier to observe. These scalar indicators are used only for interpretation, whereas the proposed LTE rainfall-intensity estimator directly processes the one-dimensional Doppler profiles.

\begin{figure*}[t]
    \centering
    \begin{subfigure}[b]{0.329\textwidth}
        \centering
        \includegraphics[width=\textwidth]{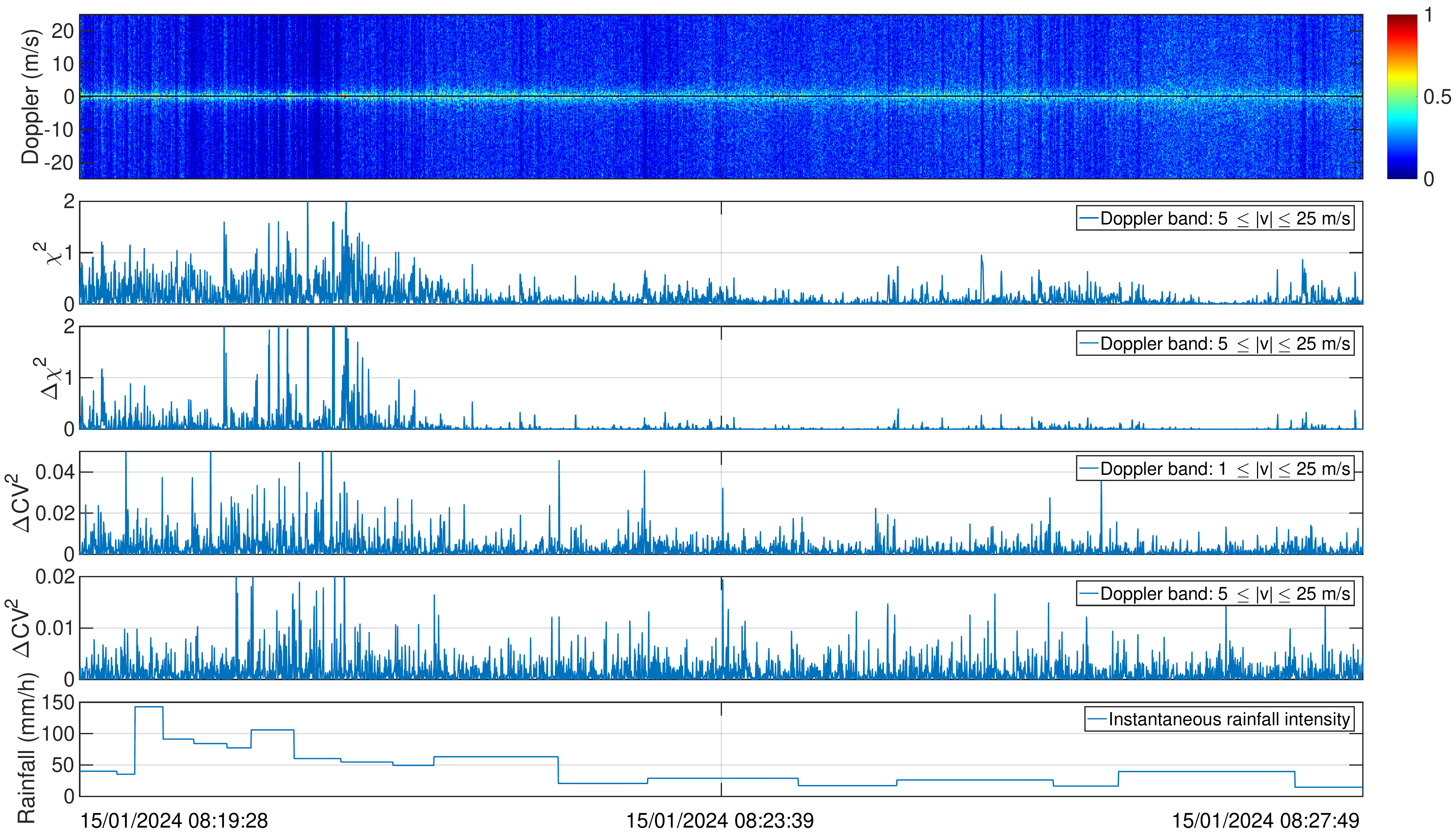}
    \end{subfigure}
    \begin{subfigure}[b]{0.329\textwidth}
        \centering
        \includegraphics[width=\textwidth]{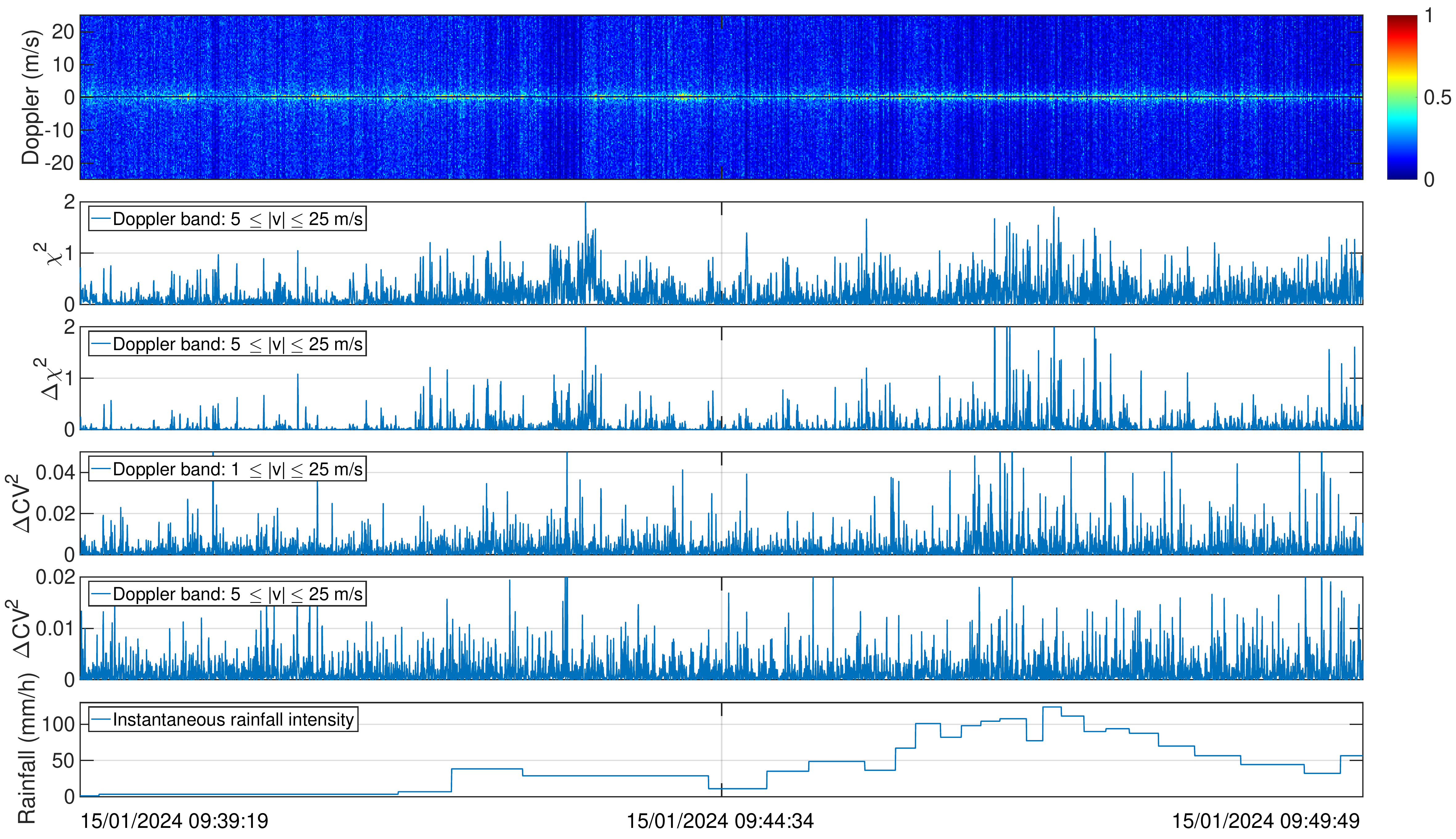}
    \end{subfigure}
    \begin{subfigure}[b]{0.329\textwidth}
        \centering
        \includegraphics[width=\textwidth]{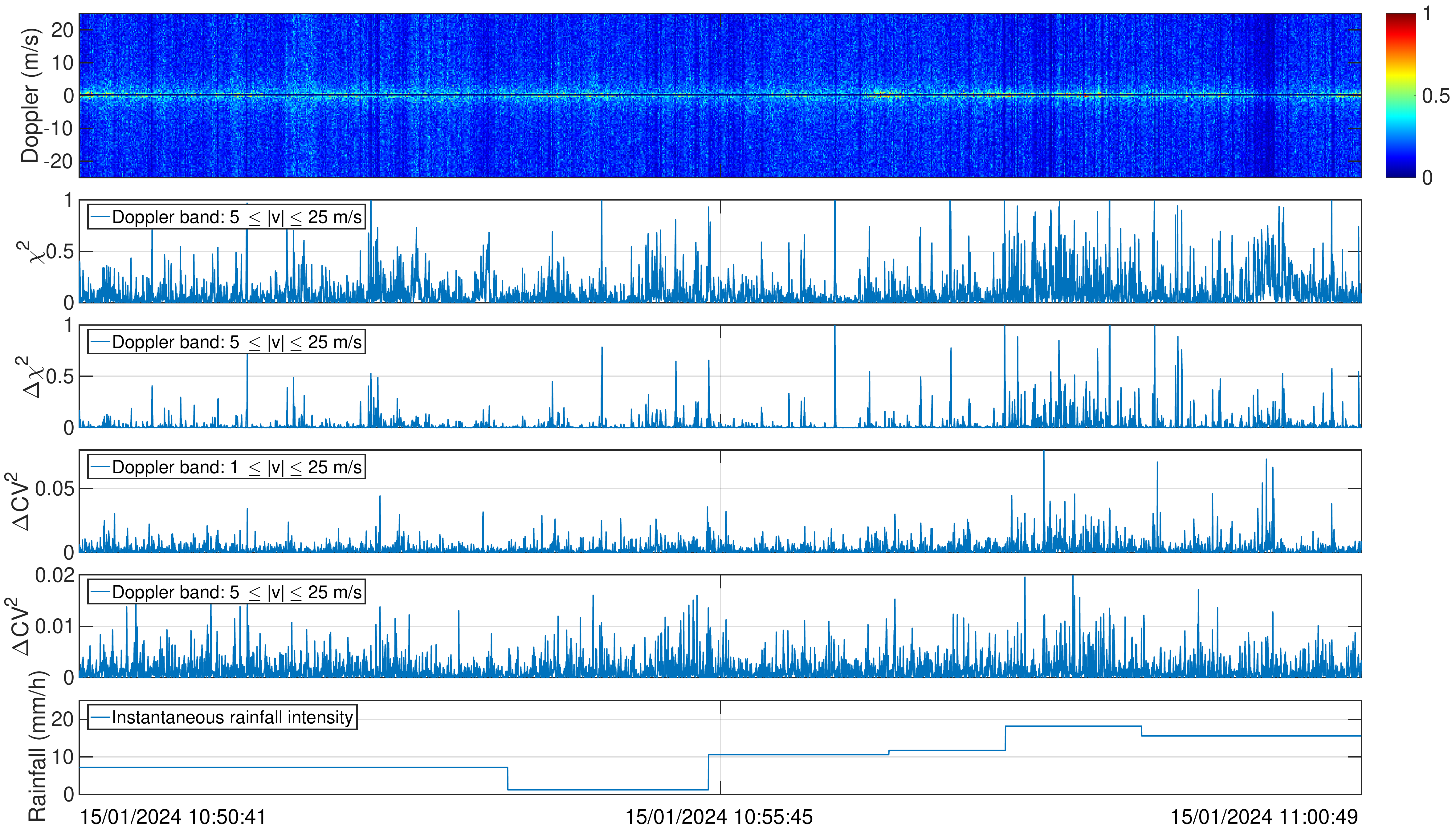}
    \end{subfigure}
    \caption{Representative results for three rainfall cases, showing the micro-Doppler, $\chi^{2}$, $\Delta\chi^{2}$, $\mathrm{CV}^{2}$, $\Delta\mathrm{CV}^{2}$, and the rainfall intensity.}
\label{fig:wifi_rainfall_features}
    \vspace{-1.5em}
\end{figure*}

\begin{figure}[t]
\centering
     \includegraphics[width=0.5\textwidth]{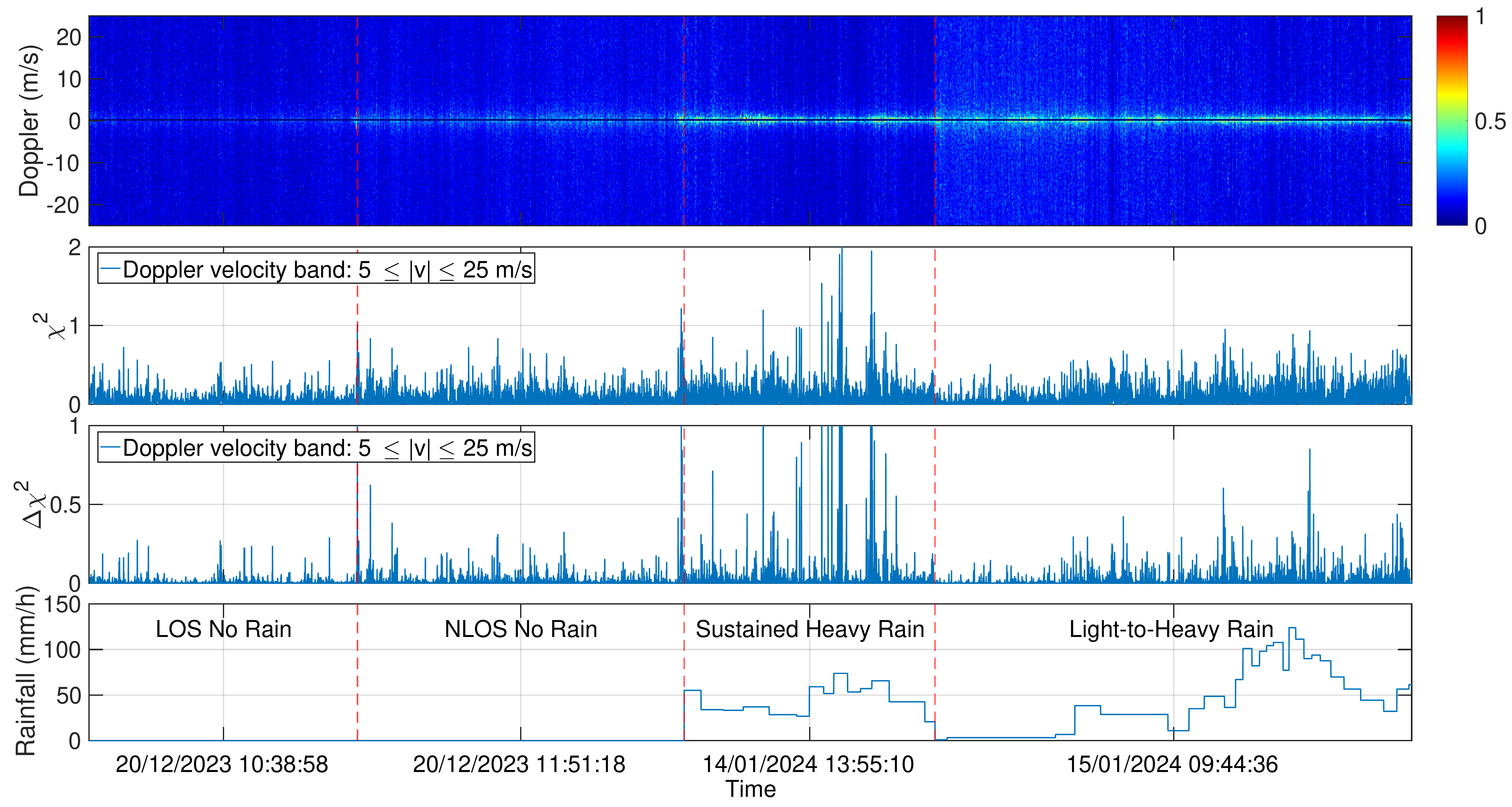}
     \caption{Rainfall features under different setups and rainfall conditions.}
     \label{fig:wifi_generalization}
\vspace{-1.5em}
\end{figure}

\section{Results}
\label{sec:experimental_results}

\subsection{WiFi Rainfall Classification Results}
\label{sec:wifi_classification_results}

\subsubsection{Representative Rainfall Feature Responses}
As shown in Fig.~\ref{fig:wifi_rainfall_features}, representative micro-Doppler spectrograms and extracted features are presented for three rainfall cases. Periods with stronger or changing rainfall generally exhibit increased Doppler fluctuations and larger $\chi^2$ and $\Delta\chi^2$ responses. While $\chi^2$ measures the deviation from the historical Doppler baseline, $\Delta\chi^2$ emphasizes short-term changes and more clearly reveals rainfall transitions. The two $\Delta\mathrm{CV}^2$ features are computed over the Doppler-velocity bands $1\leq |v|\leq25$~m/s and $5\leq |v|\leq25$~m/s, respectively, to characterize temporal changes in the Doppler-profile shape under different velocity selections. However, both show a weaker correspondence with rainfall intensity than $\Delta\chi^2$. Overall, $\Delta\chi^2$ provides the clearest temporal agreement with the ground-truth rainfall intensity.

\subsubsection{Feature Robustness Across Different Deployments}
Fig.~\ref{fig:wifi_generalization} compares the WiFi rainfall-sensitive features under LoS and NLoS deployments with no rain, sustained heavy rain, and light-to-heavy rain. Since $\chi^2$ is normalized using the historical baseline of each deployment, its absolute magnitude may vary across setups. Nevertheless, the no-rain periods remain relatively stable, whereas rainfall produces stronger and more frequent fluctuations. The $\Delta\chi^2$ feature further reduces deployment-dependent offsets and provides more consistent rainfall responses across the considered setups. These results demonstrate the robustness of the proposed feature without requiring a common absolute-power reference.

\subsubsection{Comparison with Energy-Based Features}
\begin{figure}[t]
    \centering
    \includegraphics[width=0.95\linewidth]{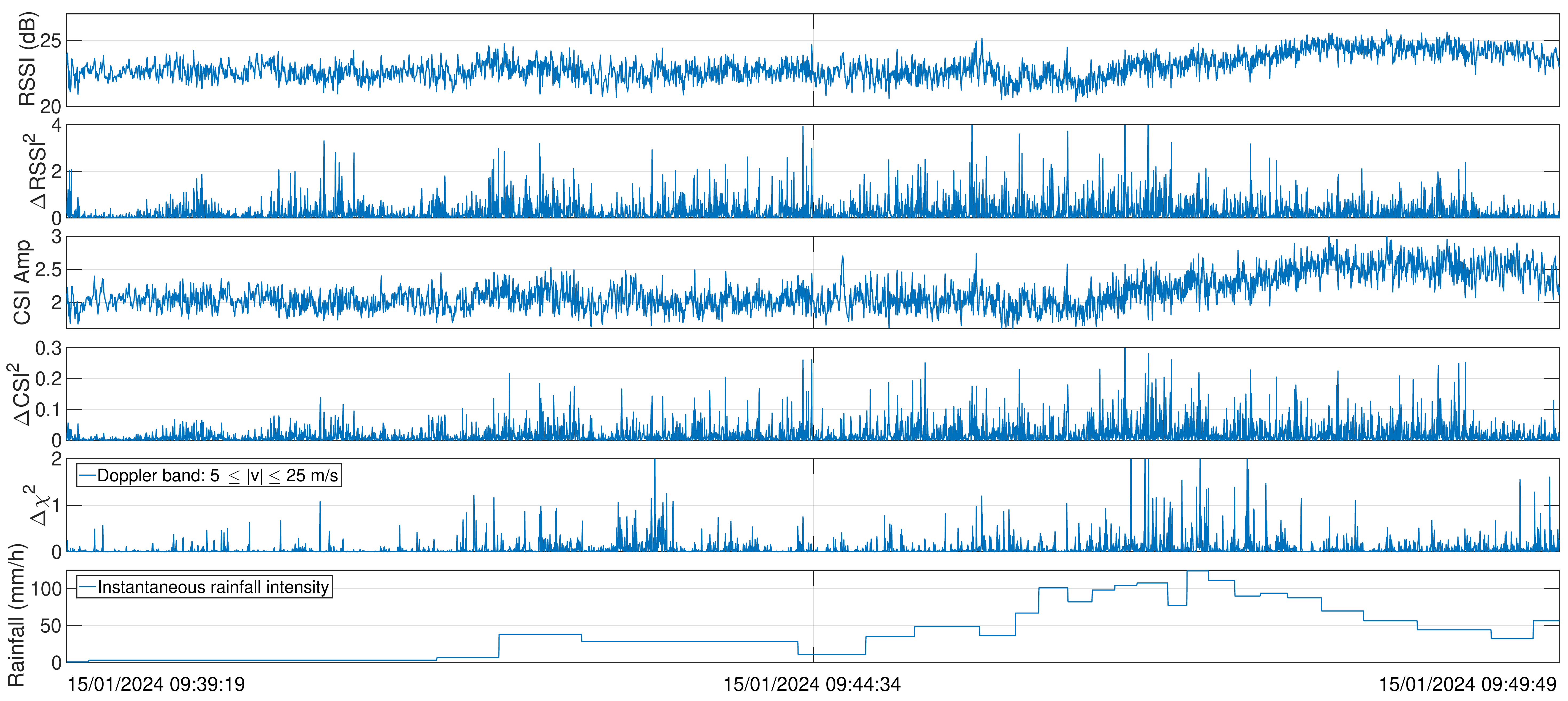}
    \caption{Feature comparison of $\Delta\chi^{2}$ with RSSI and CSI amplitude.}
    \label{fig:energy_feature_comparison}
    \vspace{-1.5em}
\end{figure}

Fig.~\ref{fig:energy_feature_comparison} compares the Doppler-based $\Delta\chi_{\kappa}^{2}$ feature with RSSI and CSI amplitude. For a fair comparison, the squared differences between adjacent samples are also computed for both energy measurements. The temporal results show that $\Delta\chi_{\kappa}^{2}$ exhibits clearer rainfall-related variations, whereas the squared differences of RSSI and CSI amplitude show only weak corresponding patterns and fluctuate during some no-rain periods. This demonstrates that the proposed Doppler feature is more sensitive to rainfall-induced channel dynamics and less affected by unrelated energy variations.

\subsubsection{Rainfall-Induced Doppler Broadening}
\begin{figure*}[t]
\centering
\begin{minipage}{0.32\textwidth}
    \centering
    \includegraphics[width=\textwidth]{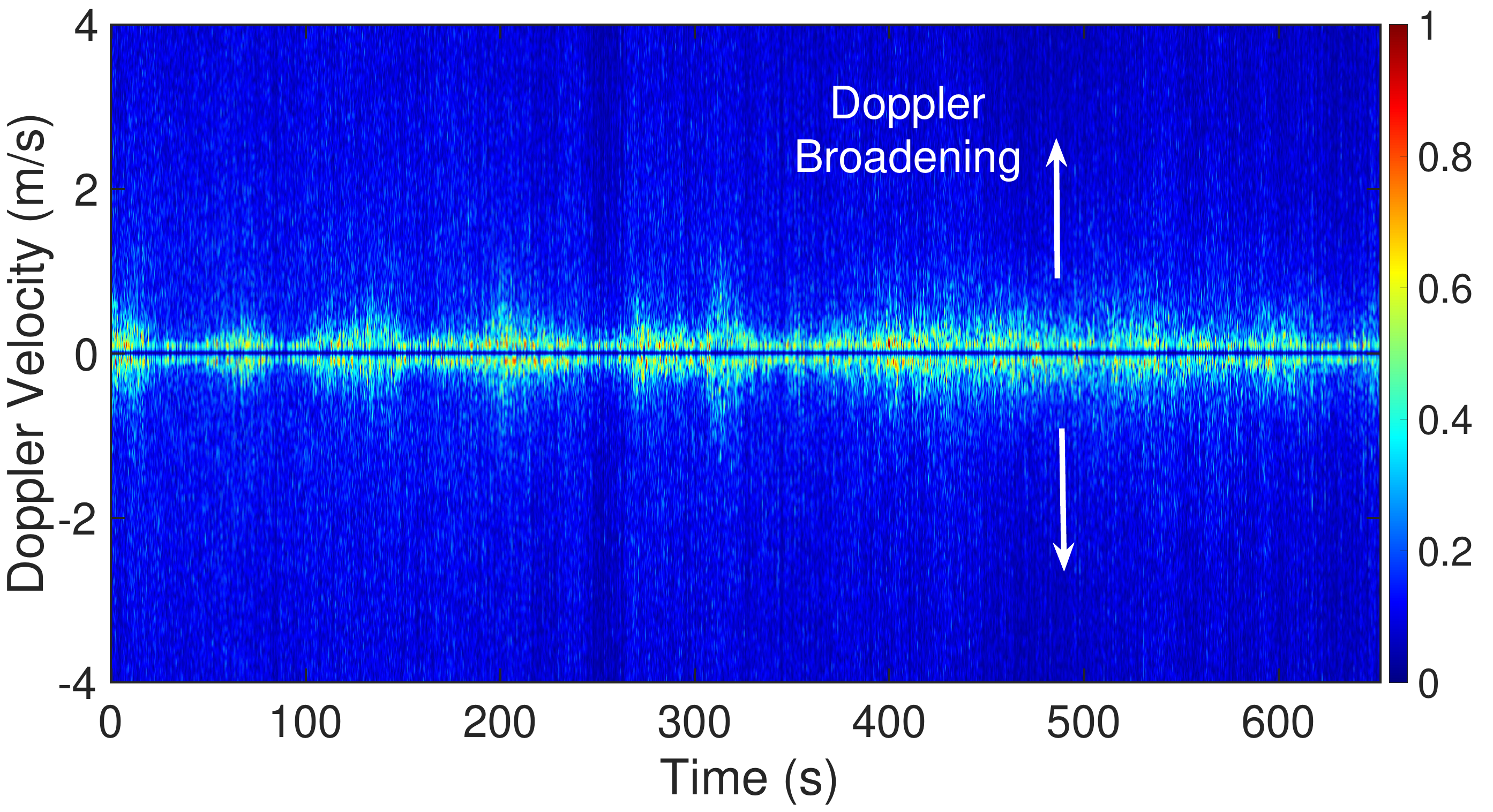}
    \caption{Doppler broadening under rainfall.}
    \label{fig:doppler_broadening}
\end{minipage}
\begin{minipage}{0.32\textwidth}
    \centering
    \includegraphics[width=\textwidth]{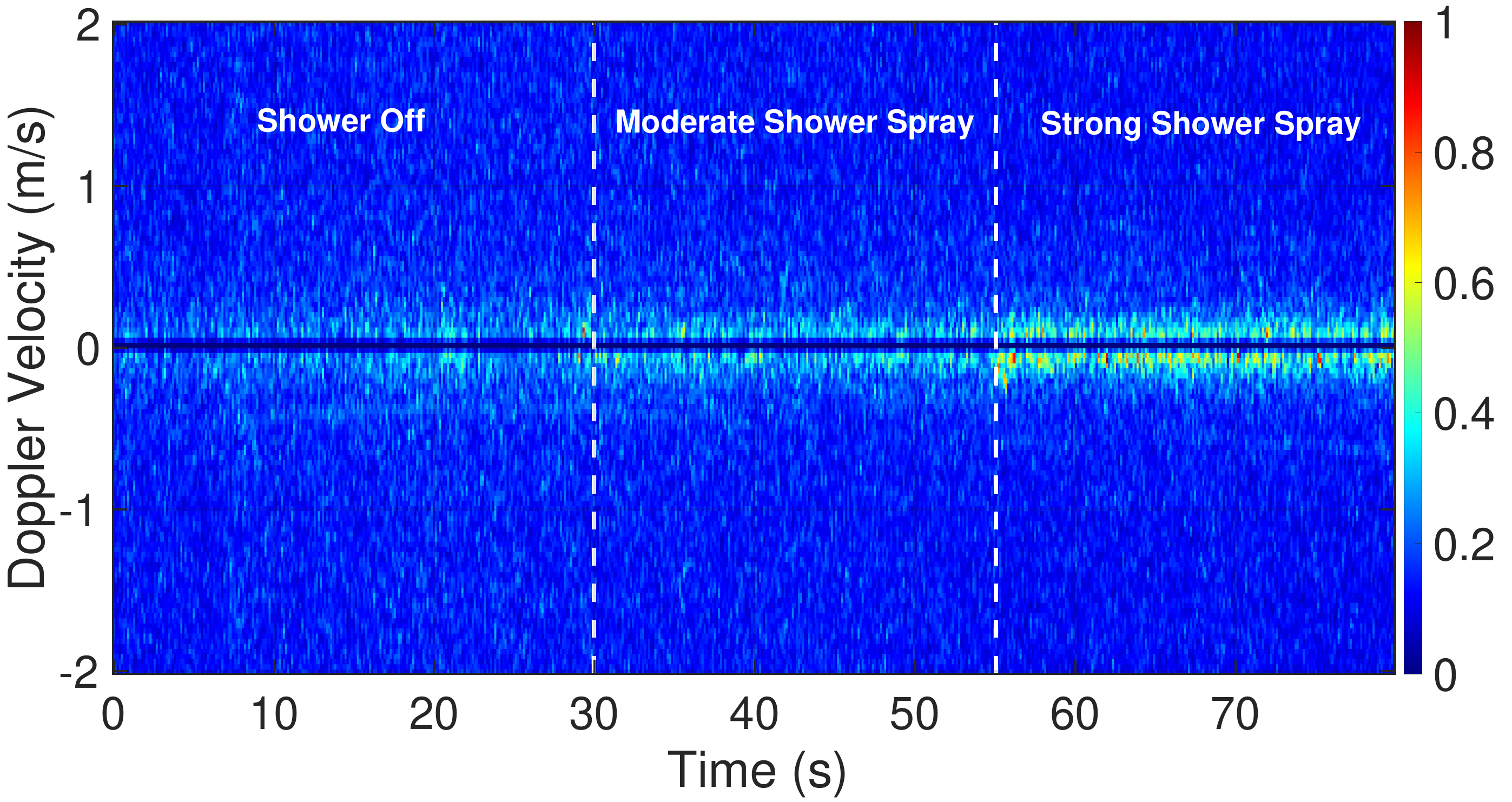}
    \caption{Controlled indoor experiment.}
    \label{fig:controlled_spray}
\end{minipage}
\begin{minipage}{0.34\textwidth}
    \centering
    \includegraphics[width=\textwidth]{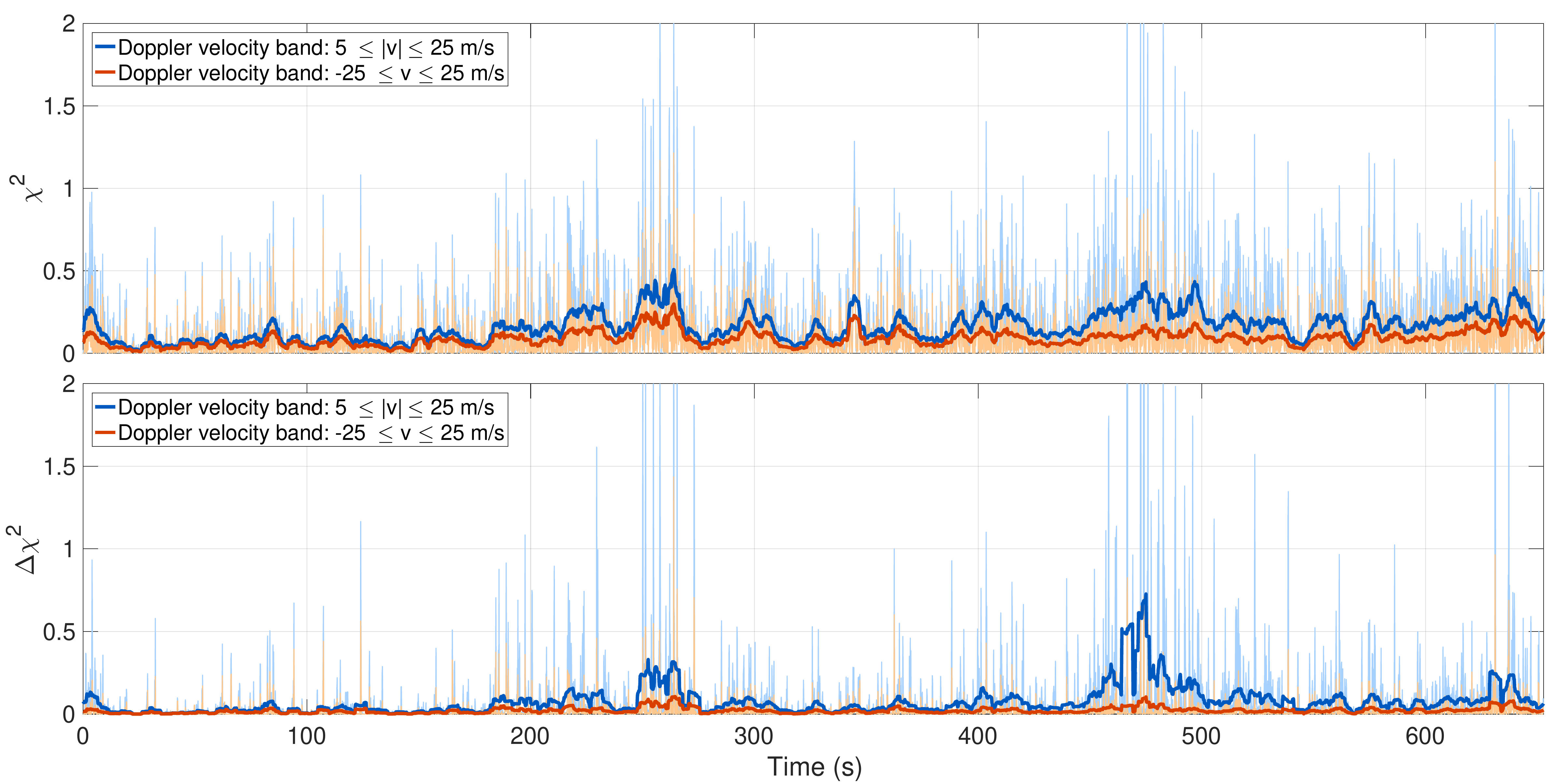}
    \caption{Doppler-band feature comparison.}
    \label{fig:doppler_comparison}
\end{minipage}
\vspace{-1em}
\end{figure*}

Fig.~\ref{fig:doppler_broadening} shows that rainfall produces increased Doppler spreading around the zero-velocity component, reflecting enhanced channel fluctuations caused by raindrop scattering. A similar trend is observed in the controlled indoor spray experiment shown in Fig.~\ref{fig:controlled_spray}, where stronger spray produces broader Doppler responses and larger feature variations. These observations demonstrate that rainfall induces measurable changes in the Doppler response.

\begin{figure*}[t]
\centering
\begin{minipage}{0.28\textwidth}
    \centering
    \includegraphics[width=0.84\textwidth]{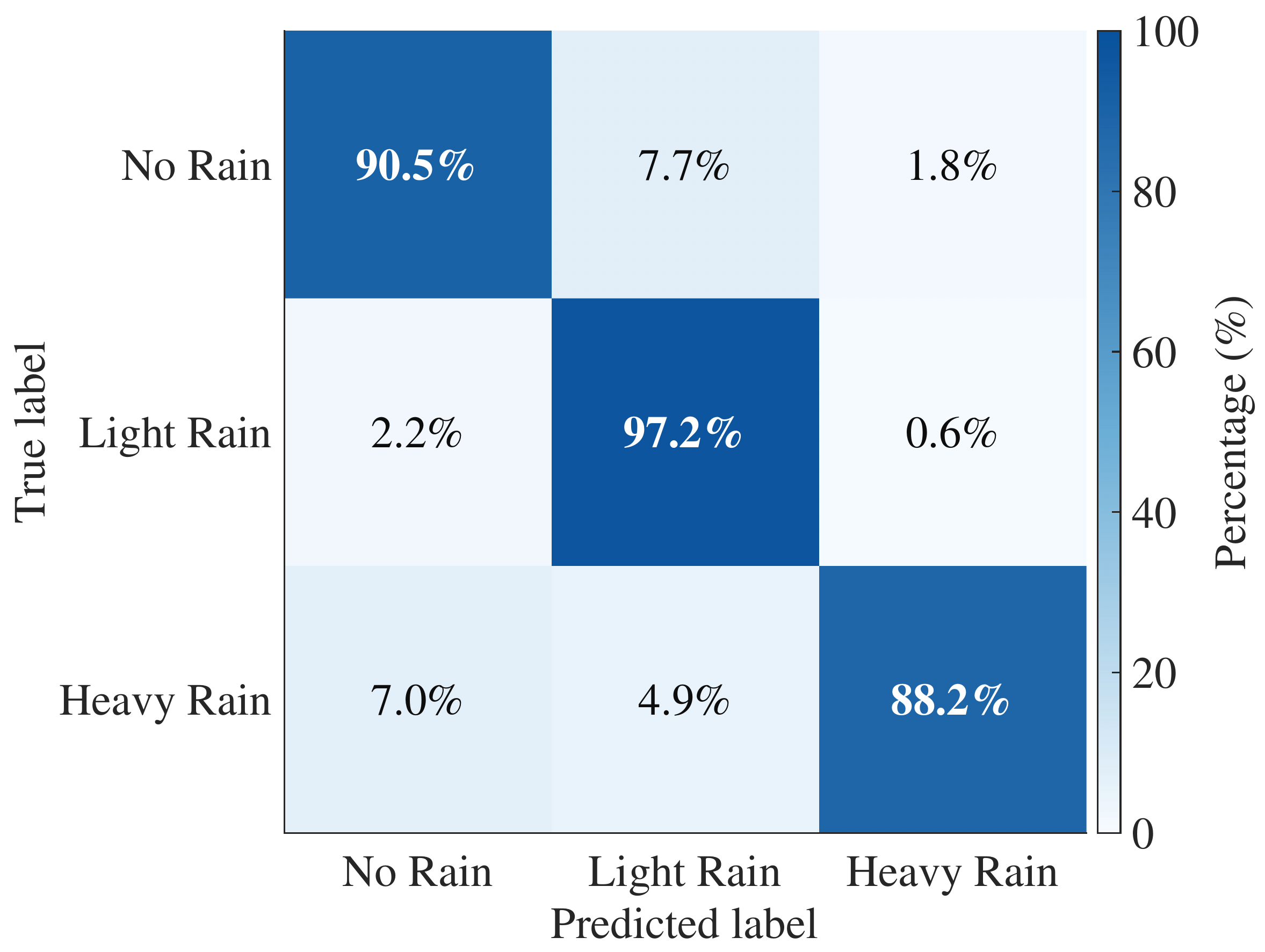}
    \caption{Confusion matrix.}
    \label{fig:confusion_matrix}
\end{minipage}
\begin{minipage}{0.355\textwidth}
    \centering
    \includegraphics[width=\textwidth]{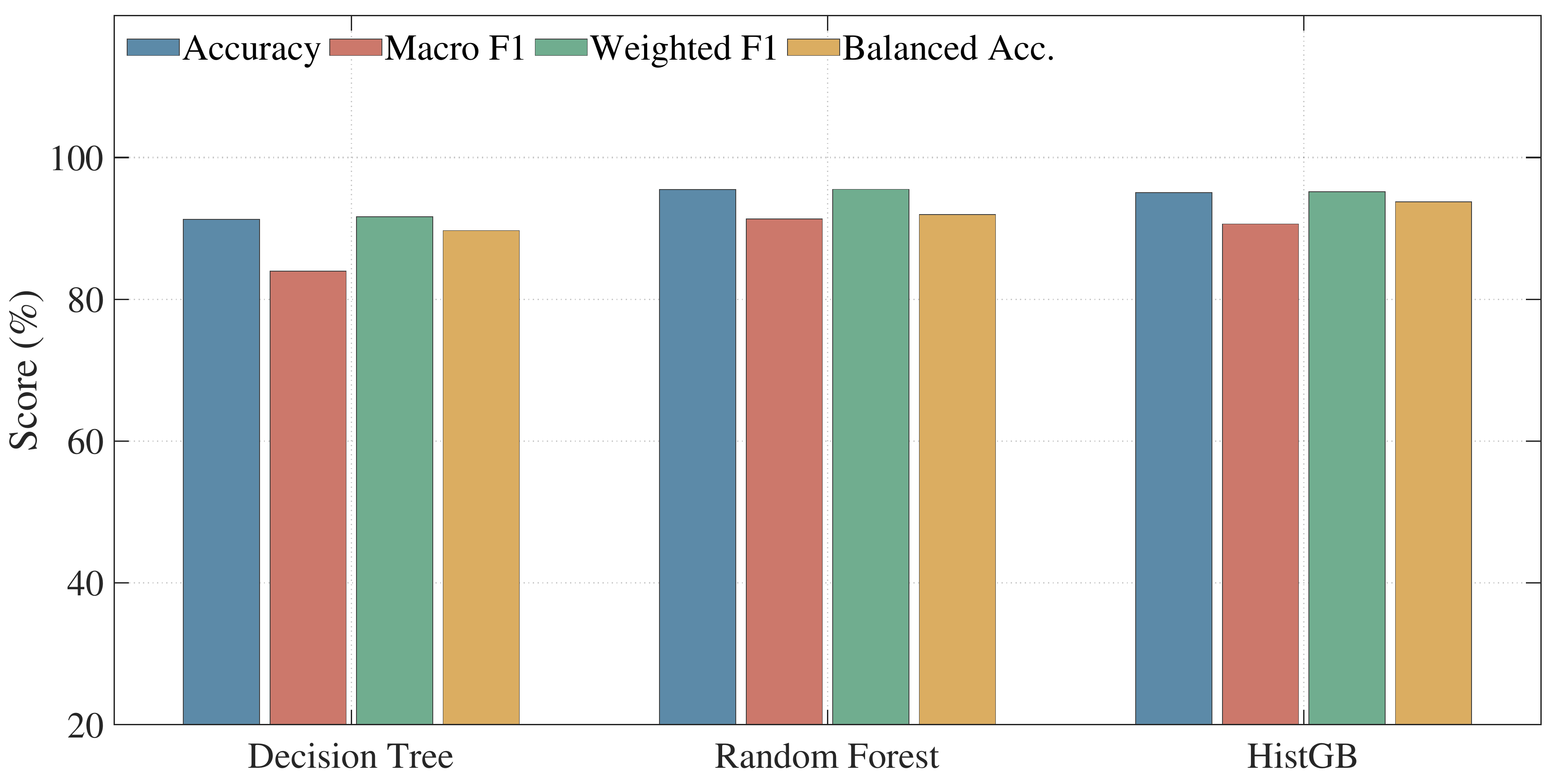}
    \caption{Baseline classifier comparison.}
    \label{fig:baseline_comparison}
\end{minipage}
\begin{minipage}{0.35\textwidth}
    \centering
    \includegraphics[width=\textwidth]{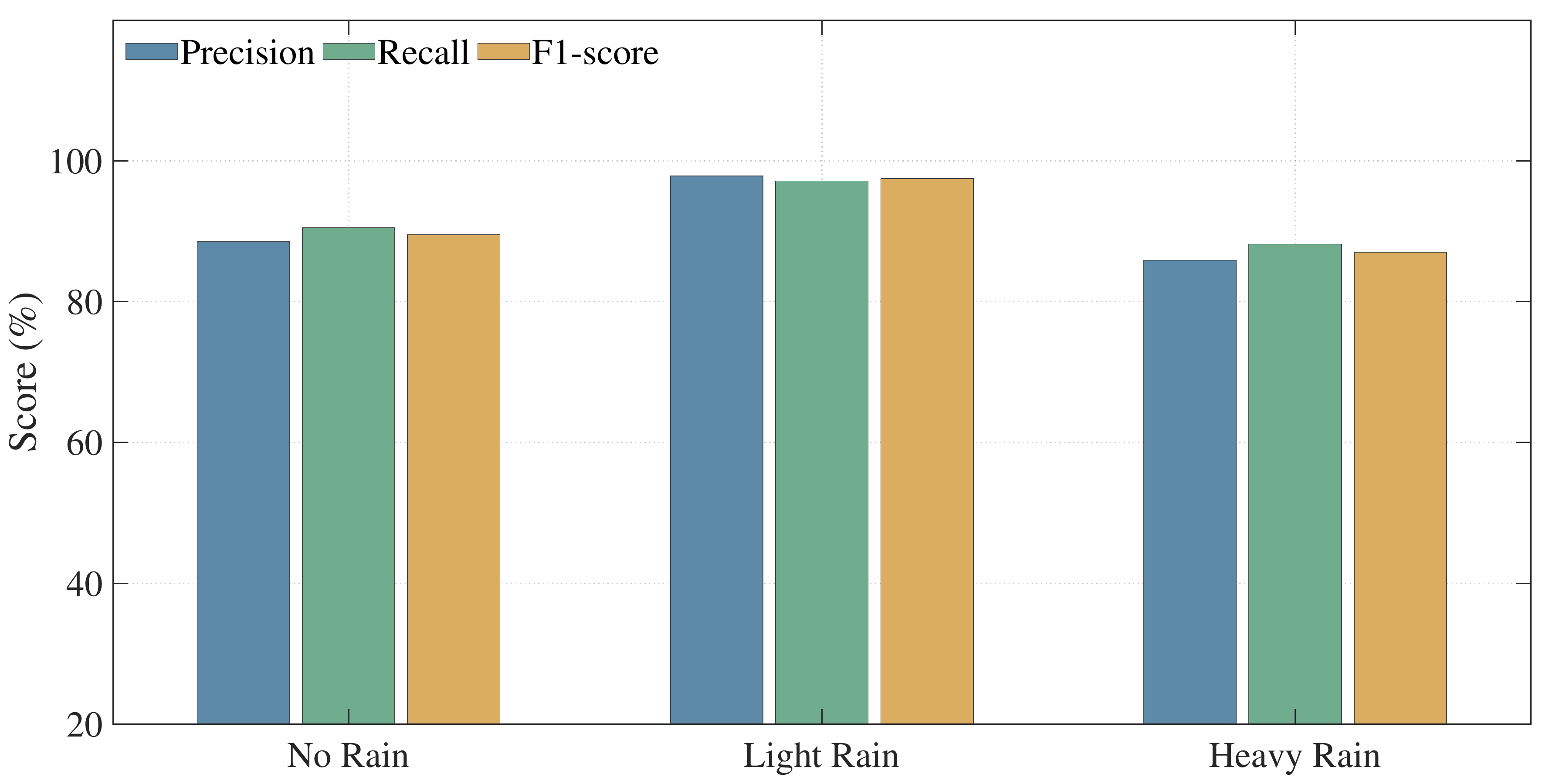}
    \caption{Per-class classification metrics.}
    \label{fig:per_class_metrics}
\end{minipage}
\vspace{-1em}
\end{figure*}

\begin{figure}[t]
\centering
\includegraphics[width=0.5\textwidth]{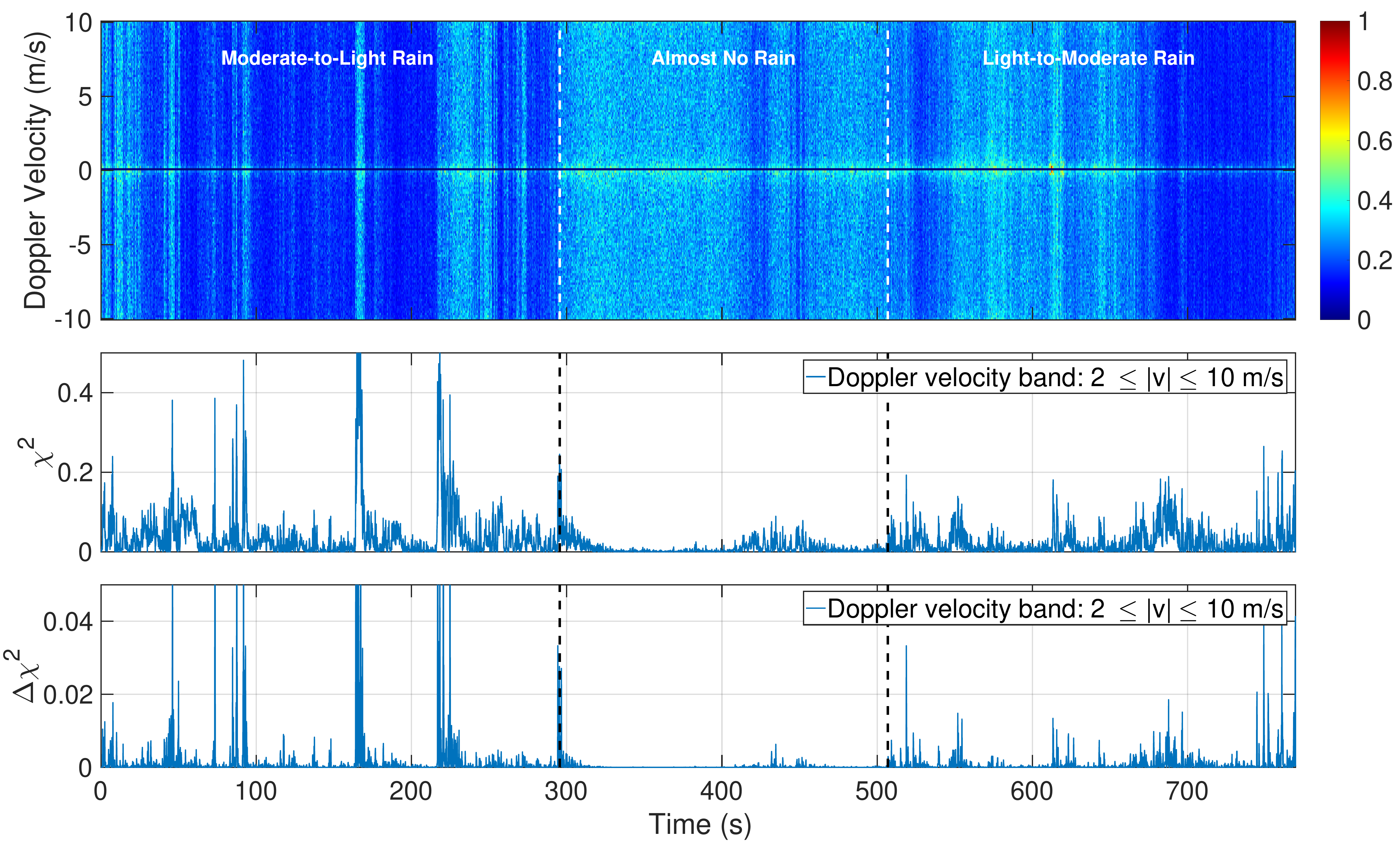}
    \caption{28 GHz mmWave rainfall Doppler response.}
    \label{fig:mmwave_outdoor_rainfall}
\vspace{-1.5em}
\end{figure}

\subsubsection{Feature Comparison across Doppler Bands}
Fig.~\ref{fig:doppler_comparison} compares the extracted features using two Doppler-velocity bands. The blue curves, obtained from the selected band $5\leq |v|\leq25$~m/s, exhibit larger $\chi^2$ and $\Delta\chi^2$ responses with clearer temporal variations. In contrast, the red curves, computed over the full band $|v|\leq25$~m/s, show weaker responses because the dominant near-zero Doppler components dilute the rainfall-induced variations. These results indicate that excluding the near-zero Doppler region suppresses background interference and improves rainfall discrimination.

\subsubsection{Rainfall Classification}
The final WiFi rainfall-classification dataset contains 155,900 samples generated from 111 measurement files. The No Rain, Light Rain, and Heavy Rain classes contain 22,063, 123,940, and 9,897 samples, respectively. To prevent leakage caused by overlapping sliding windows, the train--test split is performed at the file level, such that samples from the same file appear exclusively in either the training or testing set. The resulting split contains 111,705 training samples from 77 files and 44,195 testing samples from 34 files. Each sample is represented by five statistical features. Class-balanced sample weights are applied during training, and Random Forest \cite{scornet2026theory} is adopted as the final classifier. Fig.~\ref{fig:confusion_matrix} presents the confusion matrix on the held-out test set. The classifier achieves recalls of 90.5\%, 97.2\%, and 88.2\% for No Rain, Light Rain, and Heavy Rain, respectively. Most errors occur between No Rain and Light Rain, while confusion between No Rain and Heavy Rain is limited. Fig.~\ref{fig:baseline_comparison} compares different classifiers under the same data split. Random Forest achieves the highest accuracy of 95.48\% and weighted F1-score of 95.50\%, outperforming Decision Tree (91.28\% and 91.66\%) and HistGradientBoosting (95.05\% and 95.16\%). It also obtains a macro F1-score of 91.35\% and a balanced accuracy of 91.95\%, indicating the best performance across the three classes. Fig.~\ref{fig:per_class_metrics} reports the per-class precision, recall, and F1-score. Light Rain achieves the highest performance, with precision, recall, and F1-score of 97.85\%, 97.15\%, and 97.50\%, respectively. The No Rain and Heavy Rain classes achieve F1-scores of 89.52\% and 87.02\%, respectively. These results demonstrate that the proposed Doppler-domain features provide effective rainfall discrimination.

\subsection{Measured mmWave Rainfall Doppler Response}
Fig.~\ref{fig:mmwave_outdoor_rainfall} shows the rainfall response measured in an outdoor 28~GHz mmWave scenario. Owing to the short wavelength, raindrops interact more strongly with the signal and produce more noticeable scattering and Doppler perturbations. Compared with the near-no-rain period, moderate rainfall produces stronger Doppler variations and larger $\chi^2$ and $\Delta\chi^2$ responses, with $\Delta\chi^2$ more clearly capturing temporal transitions. These measurements provide direct evidence that rainfall induces detectable Doppler-domain signatures under realistic outdoor measurement conditions and motivate the Doppler-feature extraction and normalization used for sub-6~GHz rainfall sensing.

\subsection{LTE Rainfall Intensity Estimation}
\label{sec:lte_rainfall_estimation_results}
\begin{figure}[t]
\centering
\includegraphics[width=0.46\textwidth]{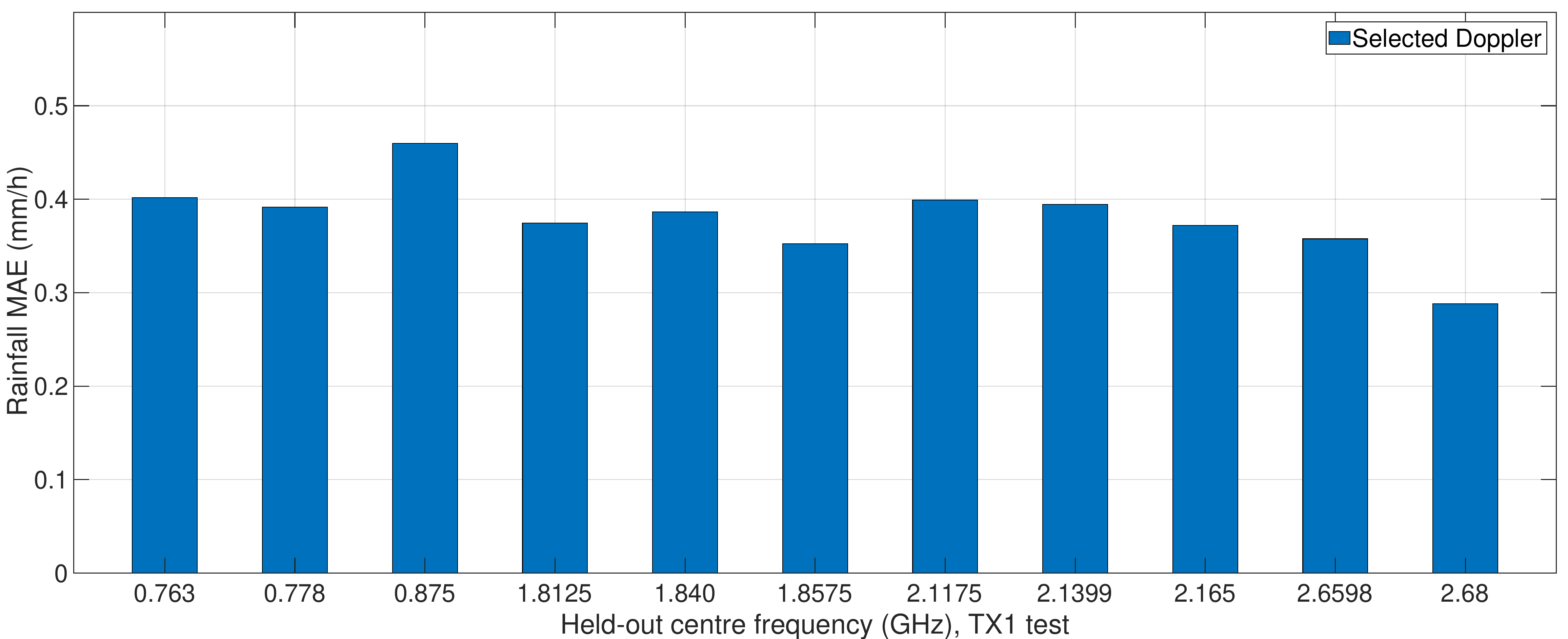}
\caption{Leave-one-frequency-out overall MAE.}
\label{fig:lte_leave_one_frequency_out}
\vspace{-1.5em}
\end{figure}

We evaluate \textit{PMN-RainSense} using the LTE dataset. Fig.~\ref{fig:lte_leave_one_frequency_out} reports leave-one-frequency-out performance, where the TX1 link at each carrier frequency is held out for testing and the remaining links are used for training. TX0 and TX1 correspond to different CRS antenna ports and are estimated from disjoint reference-signal resources, resulting in distinct transmit links with different port-specific channel responses. The overall MAE ranges from approximately 0.26 to 0.46~mm/h across the 11 carrier frequencies. Higher-frequency links generally achieve lower MAEs, with the best performance obtained at 2.68~GHz, indicating increased sensitivity to rainfall-induced channel perturbations at shorter wavelengths.

\begin{table}[t]
\centering
\small
\caption{LTE rainfall-intensity estimation results.}
\label{tab:lte_feature_ablation}
\renewcommand{\arraystretch}{1.15}
\setlength{\tabcolsep}{5pt}
\resizebox{\columnwidth}{!}{
\begin{tabular}{l|c|c|c}
\Xhline{1.2pt}
\textbf{Input features}
& \textbf{Overall MAE}
& \textbf{Rainy MAE}
& \textbf{Non-rainy MAE} \\
\Xhline{1.2pt}
Doppler (Selected)
& \textbf{0.258} & 0.706 & \textbf{0.027} \\
Doppler (Full)
& 0.277 & 0.727 & 0.046 \\
Doppler (Selected) + RSRP
& 0.260 & 0.702 & 0.034 \\
Doppler (Full) + RSRP
& 0.271 & 0.714 & 0.043 \\
RSRP
& 0.467 & \textbf{0.665} & 0.366 \\
\Xhline{1.2pt}
\end{tabular}
}
\vspace{-.5em}
\end{table}

\begin{figure*}[t]
\centering
     \includegraphics[width=\textwidth]{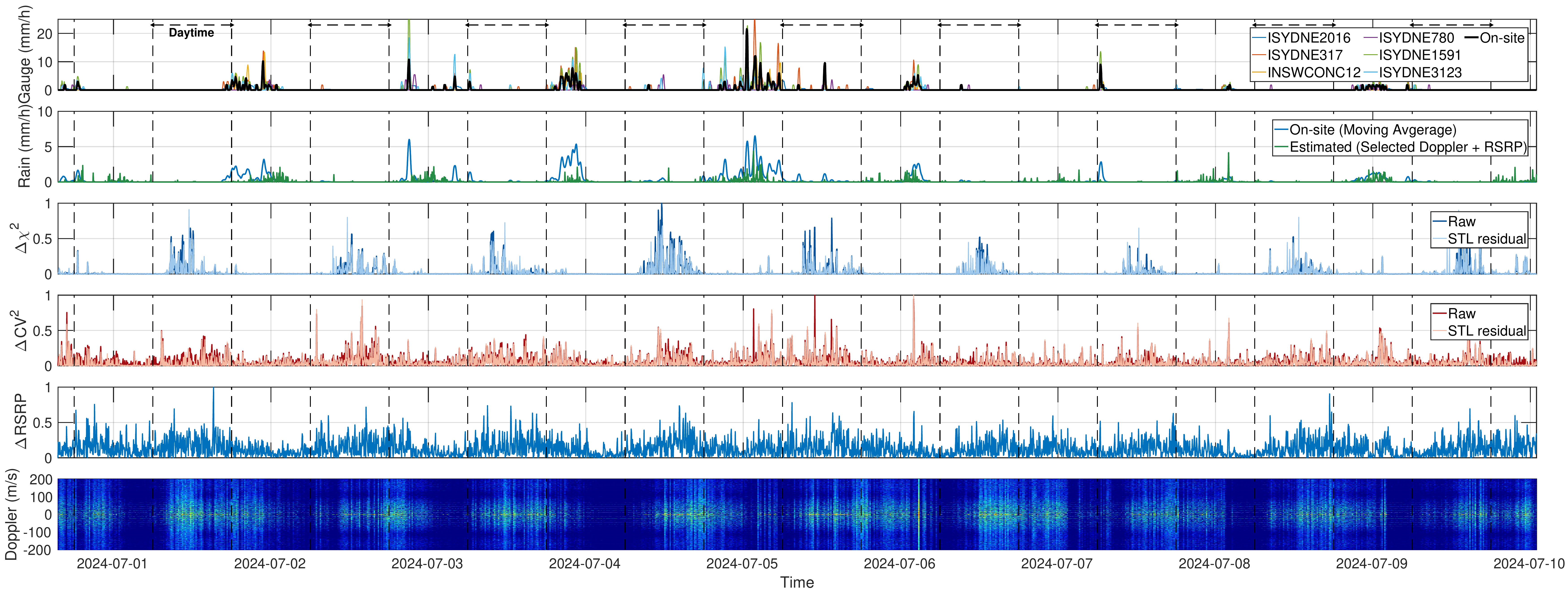}
     \caption{LTE rainfall-intensity estimation results for the held-out 2.68~GHz TX1 link from June~30 to July~10,~2024.}
     \label{fig:lte_rainfall_prediction}
\vspace{-1em}
\end{figure*}

\begin{figure*}[t]
    \centering
    \begin{subfigure}[t]{0.32\textwidth}
        \centering
        \includegraphics[width=\textwidth]{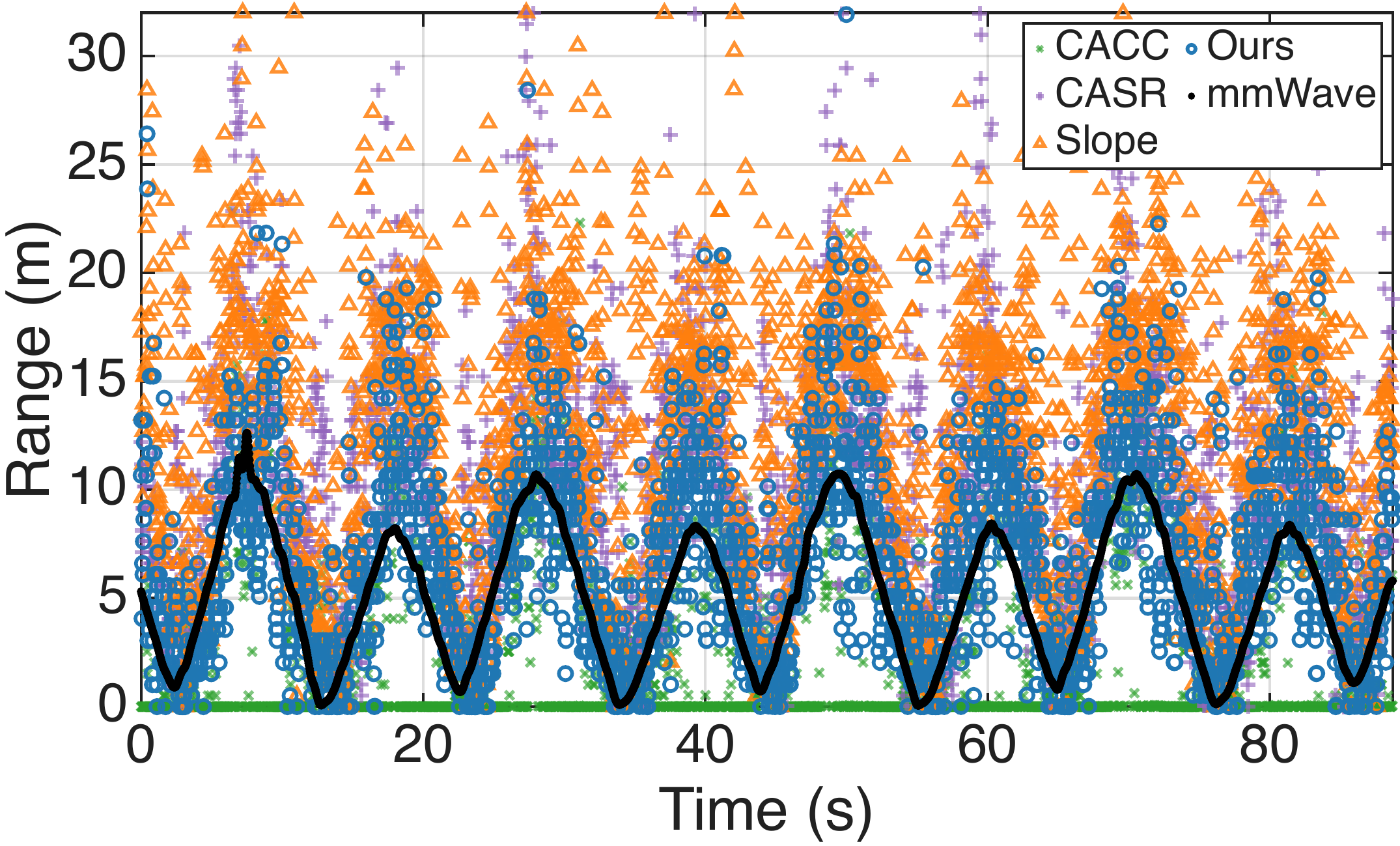}
    \end{subfigure}
    \hfill
    \begin{subfigure}[t]{0.32\textwidth}
        \centering
        \includegraphics[width=\textwidth]{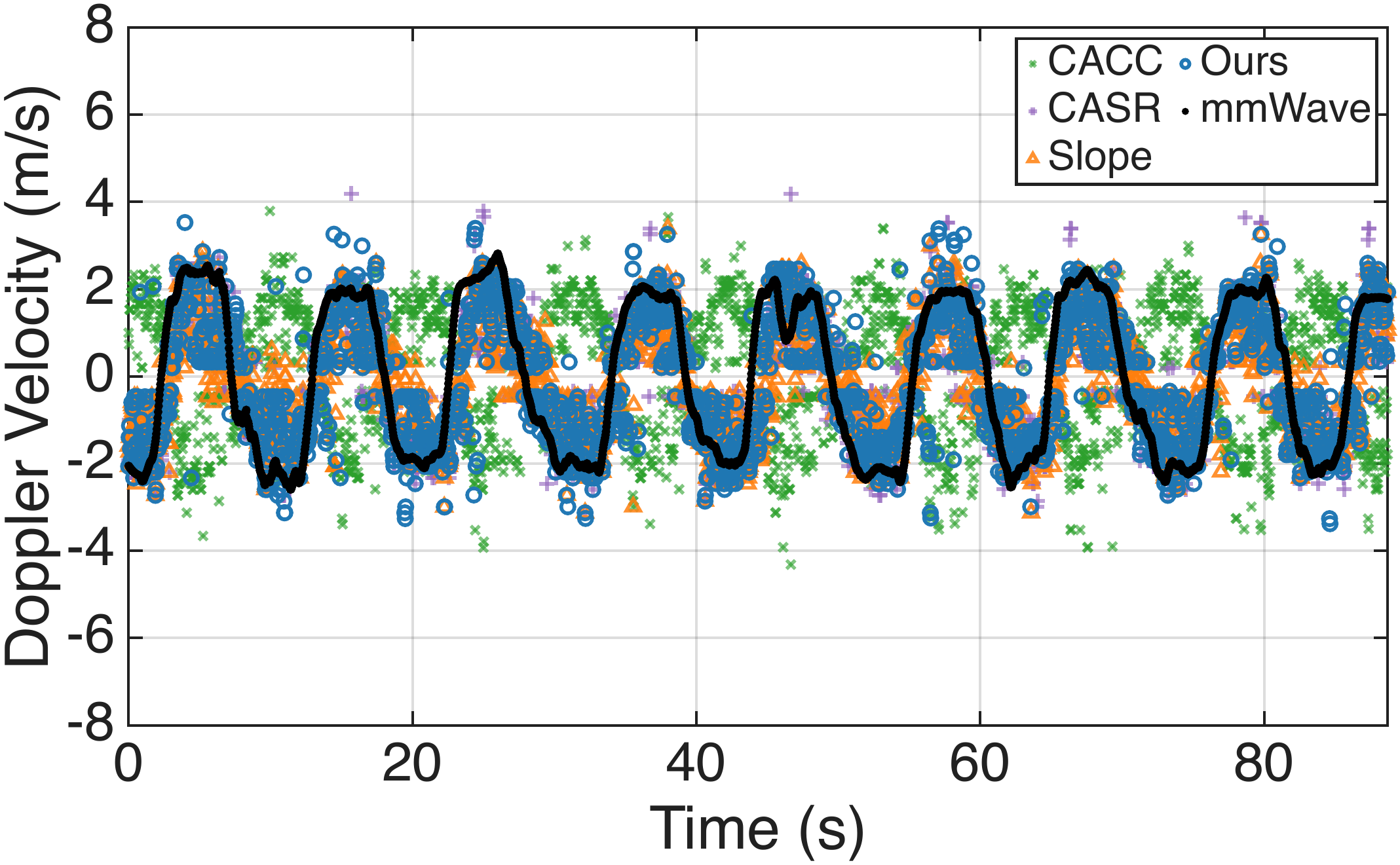}
    \end{subfigure}
    \hfill
    \begin{subfigure}[t]{0.32\textwidth}
        \centering
        \includegraphics[width=\textwidth]{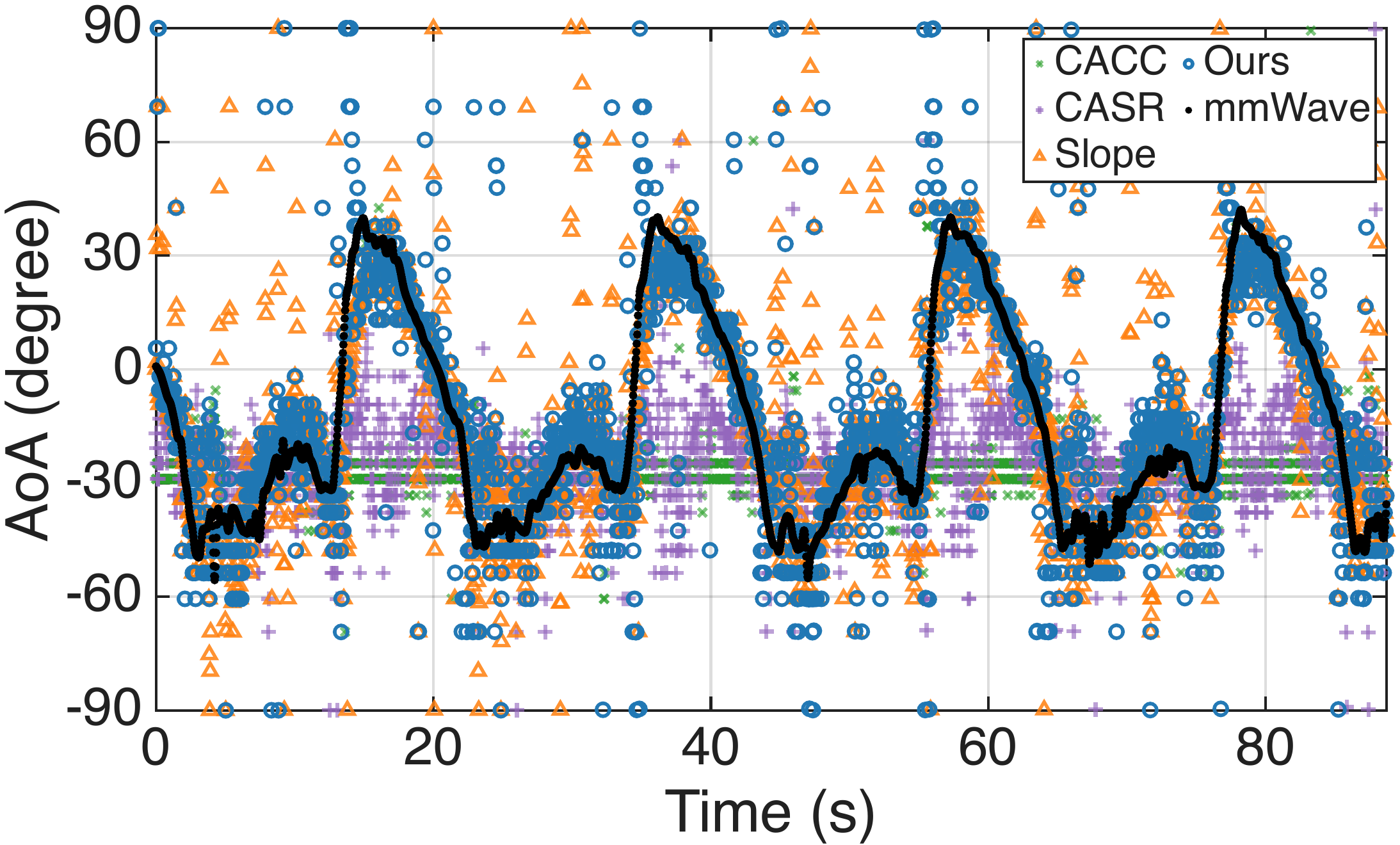}
    \end{subfigure}
    \caption{Validation of delay, Doppler, and AoA feature extraction accuracy under different CSI compensation methods.}
    \label{fig:range_doppler_aoa_comparison_ellipse}
    \vspace{-1.5em}
\end{figure*}

\begin{figure}[t]
    \centering
    \begin{subfigure}[t]{0.48\columnwidth}
        \centering
        \includegraphics[width=\linewidth]{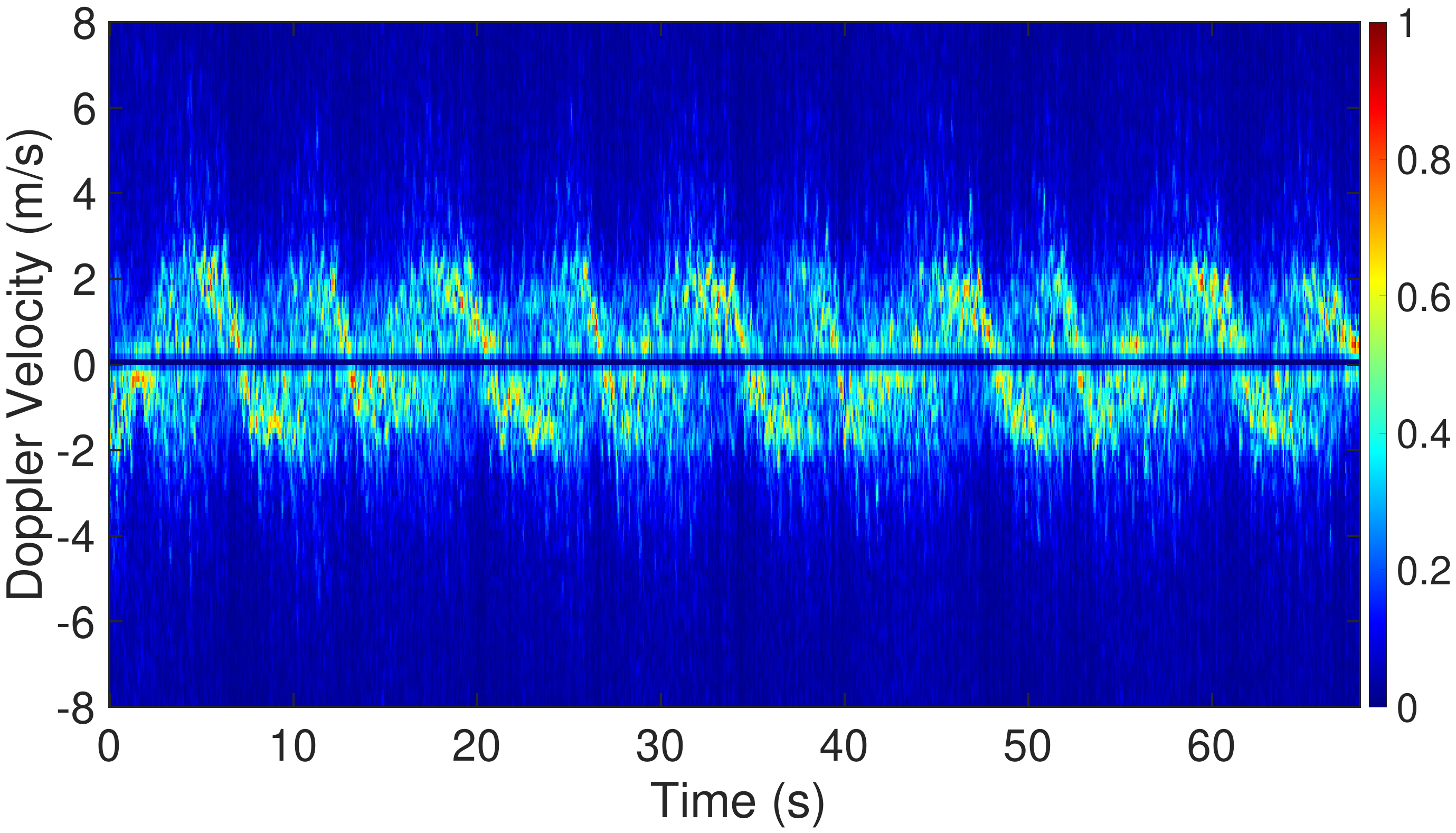}
        \caption{Micro-Doppler.}
        \label{fig:multi_target_doppler_result}
    \end{subfigure}
    \begin{subfigure}[t]{0.48\columnwidth}
        \centering
        \includegraphics[width=\linewidth]{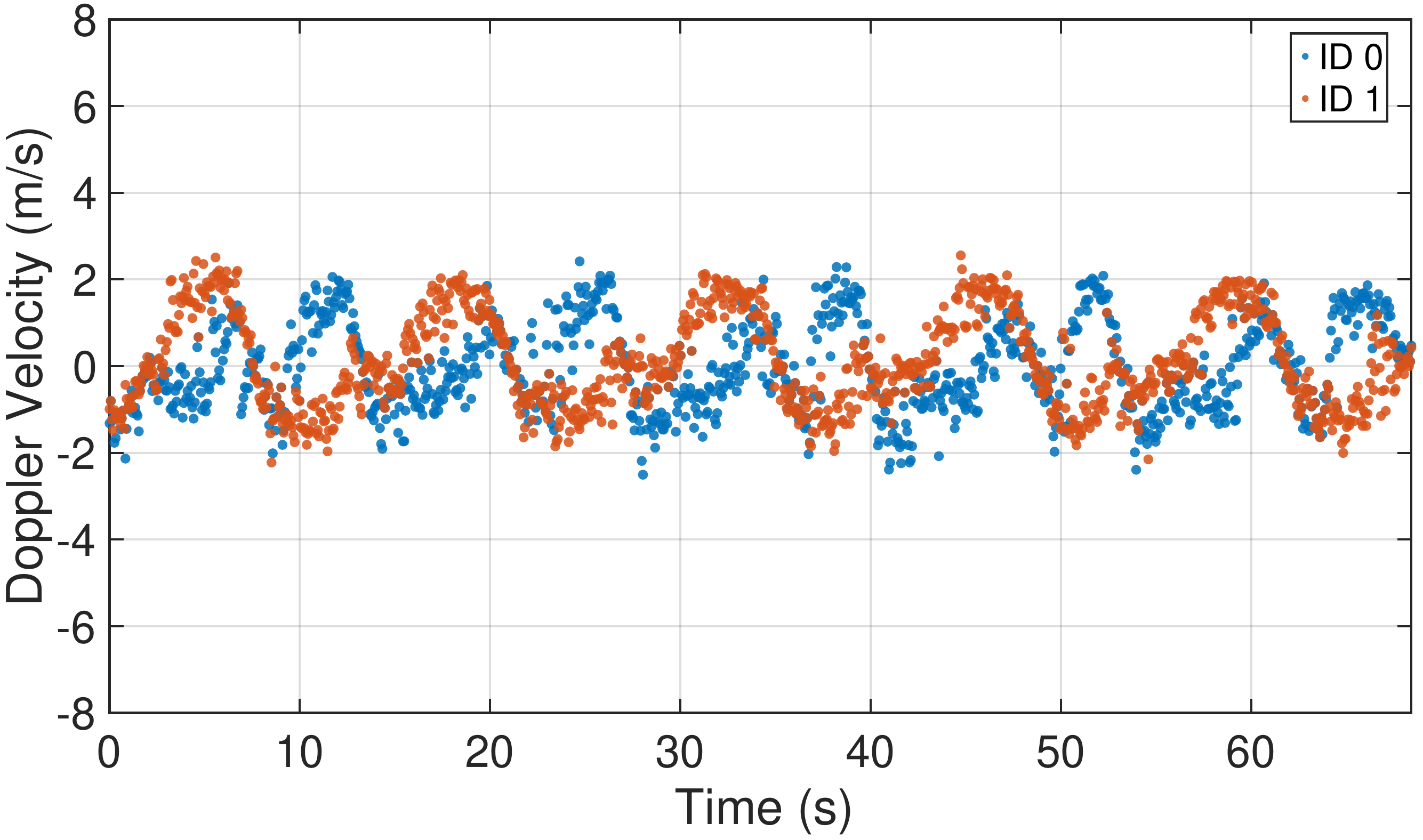}
        \caption{Target Doppler traces.}
        \label{fig:multi_target_doppler_ref}
    \end{subfigure}
    \caption{Multi-target Doppler extraction results.}
    \label{fig:multi_target_doppler_comparison}
    \vspace{-1.5em}
\end{figure}

Table~\ref{tab:lte_feature_ablation} reports the results using the 2.68~GHz TX1 link as the held-out test link and compares the selected Doppler profile, the full Doppler profile, their combinations with RSRP, and the RSRP-only baseline. The selected Doppler profile achieves the lowest overall and non-rainy MAEs of 0.258 and 0.027~mm/h, respectively. Adding RSRP slightly improves the rainy-period MAE while causing only a marginal increase in the overall MAE, indicating that RSRP provides complementary information for fitting rainfall intensity. The full Doppler profile retains near-zero components that increase the overall and non-rainy errors, whereas the RSRP-only baseline achieves the lowest rainy MAE of 0.665~mm/h but substantially higher overall and non-rainy MAEs of 0.467 and 0.366~mm/h. Overall, the selected Doppler profile provides the best robustness, while its combination with RSRP offers a favorable tradeoff for tracking rainfall intensity.

Fig.~\ref{fig:lte_rainfall_prediction} presents the results from June~30 to July~10,~2024 using the 2.68~GHz TX1 link as the held-out test link. The first panel shows rainfall measurements from six nearby public weather stations and the on-site weather station co-located with the LTE receiver, while the second compares the temporally averaged on-site rainfall labels with the estimated intensity. The third and fourth panels present the raw and STL-processed $\Delta\chi_{\kappa}^{2}$ and $\Delta\mathrm{CV}_{\kappa}^{2}$, respectively, the fifth panel shows the absolute difference between adjacent RSRP measurements, and the bottom panel displays the micro-Doppler spectrogram. Because $\Delta\chi_{\kappa}^{2}$ spans a large dynamic range, it is logarithmically transformed before visualization and STL processing. The micro-Doppler spectrogram retains all bins within $[-200,200]$~m/s, whereas $\Delta\chi_{\kappa}^{2}$ and $\Delta\mathrm{CV}_{\kappa}^{2}$ are derived from the selected Doppler profile satisfying $20\leq |v|\leq200$~m/s to suppress near-zero background fluctuations. STL \cite{cleveland1990stl} is applied separately to the two indicator sequences using a period of 480 samples, corresponding to a 24-hour cycle at the approximately 3-minute measurement interval. As shown in the third and fourth panels, STL attenuates the dominant diurnal pattern, although residual variations remain because environmental and network activities do not repeat identically each day. Compared with the absolute difference between adjacent RSRP measurements, the Doppler-derived indicators exhibit clearer responses during several rainfall periods. Overall, dynamic environmental interference remains a major challenge, and future work will investigate larger datasets and more advanced interference-suppression methods.

\subsection{Validation of Delay, Doppler, and AoA Feature Extraction}
This subsection validates whether the proposed CSI compensation and feature extraction method preserves physically meaningful delay, Doppler, and AoA information. Fig.~\ref{fig:range_doppler_aoa_comparison_ellipse} compares the estimated parameters for a single-target elliptical trajectory with mmWave radar measurements as the reference. The proposed method is evaluated against several representative compensation schemes, including Slope \cite{11431712}, CACC \cite{zhang2021widar3}, and CASR \cite{zeng2019farsense}. Residual TO after Slope compensation introduces a noticeable bias in the range estimates. The conjugate multiplication adopted by CACC produces mirror Doppler components, making the target motion direction ambiguous, while the nonlinear ratio operation in CASR degrades the delay and AoA estimates. In contrast, our method follows the reference trajectory more consistently across all three domains. Fig.~\ref{fig:multi_target_doppler_comparison} further demonstrates that, for two targets moving in opposite directions, the compensated CSI preserves both superimposed Doppler components. These results validate the effectiveness of our CSI compensation and feature extraction pipeline and support its application to rainfall sensing.

\section{Conclusion}
This paper proposes \textit{PMN-RainSense}, a passive rainfall-sensing framework using wireless communication signals. The proposed CSI compensation and Doppler-feature extraction method suppresses hardware impairments while preserving rainfall-sensitive channel variations. By exploiting delay--Doppler representations and physics-guided Doppler filtering, the framework improves robustness to background interference and link-dependent variations. Experiments using WiFi, LTE, and mmWave measurements consistently verify rainfall-induced Doppler responses, achieving 95.48\% WiFi classification accuracy and an LTE rainfall-intensity MAE of 0.258~mm/h on a held-out transmit-port link. The mmWave measurements further confirm rainfall-associated Doppler broadening at higher carrier frequencies. These results demonstrate the feasibility of passive rainfall monitoring using existing wireless infrastructure and provide a basis for dense and continuous environmental sensing.

\bibliographystyle{IEEEtran}
\bibliography{main.bib}

\end{document}